\pdfoutput=1

\documentclass[fleqn,usenatbib]{mnras}

\usepackage{newtxtext,newtxmath}

\usepackage{bm}
\usepackage[T1]{fontenc}
\usepackage{natbib,float}
\usepackage{ulem,stfloats}
\DeclareRobustCommand{\VAN}[3]{#2}
\let\VANthebibliography\thebibliography
\def\thebibliography{\DeclareRobustCommand{\VAN}[3]{##3}\VANthebibliography}

\makeatletter

\newcommand{\Rmnum}[1]{\expandafter\@slowromancap\romannumeral #1@}
\makeatother

\usepackage{siunitx}

\usepackage{graphicx}	
\usepackage{amsmath}	
\usepackage{physics}

\makeatletter
\newlength{\abovecaptionskip}%
\makeatother
\usepackage{threeparttable}

\newcommand{\xmm}{{\it XMM-Newton}}

\newcommand{\msun}{\ensuremath{M_{\sun}}}
\newcommand{\mbh}{\ensuremath{M_{\mathrm{BH}}}}
\newcommand{\ledd}{\ensuremath{L_{\mathrm{Edd}}}}
\newcommand{\leddrat}{\ensuremath{\lambda_{\mathrm{Edd}}}}
\newcommand{\lumcgs}{\ensuremath{\mathrm{erg}\,\mathrm{s}^{-1}}}
\newcommand{\fluxcgs}{\ensuremath{\mathrm{erg}\,\mathrm{s}^{-1}\,\mathrm{cm}^{-2}}}
\newcommand{\fluxdenscgs}{\ensuremath{\mathrm{erg}\,\mathrm{s}^{-1}\,\mathrm{cm}^{-2}\,\mathrm{Hz}^{-1}}}

\newcommand{\sx}[1]{\ensuremath{\textit{SX}#1}}
\newcommand{\kbol}{\ensuremath{k_\mathrm{bol}}}
\newcommand{\Lbol}{\ensuremath{L_\mathrm{bol}}}

\title[\xmm\ study on six high-z NLS1s]{An \xmm\ Study of Six Narrow-Line Seyfert 1 Galaxies at $z=0.35-0.92$}

\author[Zhibo Yu et al.]{
Zhibo Yu,$^{1,2,3}$\thanks{E-mail: zvy5225@psu.edu}
Jiachen Jiang,$^{4}$
Cosimo Bambi,$^{3}$ Luigi C. Gallo,$^{5}$ Dirk Grupe,$^{6}$ Andrew C. Fabian,$^{4}$ \newauthor
Christopher S. Reynolds,$^{4}$ William N. Brandt$^{1,2,7}$
\\
$^{1}$Department of Astronomy and Astrophysics, The Pennsylvania State University, 525 Davey Lab, University Park, PA 16802, USA\\
$^{2}$Institute for Gravitation and the Cosmos, The Pennsylvania State University, University Park, PA 16802, USA\\
$^{3}$Center for Field Theory and Particle Physics and Department of Physics, Fudan University, 2005 Songhu Road, Shanghai 200438, China\\
$^{4}$Institute of Astronomy, University of Cambridge, Madingley Road, Cambridge CB3 0HA, UK\\
$^{5}$Department of Astronomy and Physics, Saint Mary’s University, 923 Robie Street, Halifax, NS, B3H 3C3, Canada\\
$^{6}$Department of Physics, Geology, and Engineering
Technology, Northern Kentucky University, Nunn Drive, Highland Heights, KY 41099, USA\\
$^{7}$Department of Physics, The Pennsylvania State University, 104 Davey Laboratory, University Park, PA 16802, USA
}

\date{Accepted XXX. Received YYY; in original form ZZZ}

\pubyear{2022}

\begin{document}
\label{firstpage}
\pagerange{\pageref{firstpage}--\pageref{lastpage}}
\maketitle

\begin{abstract}
    We report a detailed analysis of the \xmm\ spectra of six Narrow-Line Seyfert 1 (NLS1) galaxies at redshift $z=0.35-0.92$. Compared with the NLS1s at lower redshift in the previously most-studied sample, these NLS1s have larger black hole (BH) masses ($\log\,M_\mathrm{BH}>7.5$) with similar or even lower Eddington ratios. Our extended \xmm\ sample of NLS1s shows strong soft X-ray excess emission below 2 keV. The quantified soft excess strength does not show an obvious discrepancy from previous studies of the lower-redshift NLS1s. The systematic effect in the measurement of the Eddington ratio mainly lies in the bolometric correction factor. We also tentatively fit the spectra assuming two more physical models for the soft excess: warm Comptonization and relativistic reflection from the inner accretion disk. In the first scenario, we confirm the ubiquity of a warm and optically thick corona. The behavior of a single source can be better explained by relativistic reflection, although we cannot distinguish which model is a more favorable explanation for the soft excess based on the best-fit statistics.
\end{abstract}




\section{Introduction}

\subsection{Seyfert 1 galaxies and NLS1s}

Active Galactic Nuclei (AGNs) radiate across the electromagnetic spectrum from radio to $\gamma$-ray, often even brighter than the rest of their host galaxies \citep{2002apa..book.....F}. Seyfert galaxies are one class of active galaxies with luminous nuclei \citep{peterson1997introduction}. Depending upon the width of the emission lines in the spectrum, Seyfert galaxies are classified into Seyfert 1 and 2 galaxies. Seyfert 1 galaxies are characterized by broad (corresponding to a velocity of $\approx10^4\,\si{km.s^{-1}}$) permitted lines and narrow forbidden lines, while the permitted lines in Seyfert 2 galaxies  are similar but narrower ($\lesssim10^3\,\si{km.s^{-1}}$) \citep{karttunen2007fundamental}. Narrow-line Seyfert 1 (NLS1) galaxies are a subclass of Seyfert 1 introduced in \citet{1985ApJ...297..166O}. They show similar spectral properties, but the line widths are small compared with their broad-line counterparts. NLS1s are defined by a narrow permitted H$\beta$ line with full width at half maximum (FWHM) $<$ \SI{2000}{\km\per\second}, [\ion{O}{iii}]/H$\beta$ $<$ 3 \citep{1983ApJ...273..478O,1989ApJ...342..224G} and often strong \ion{Fe}{ii} emission.

NLS1s are believed to host relatively lower-mass supermassive black holes (BHs) with $M_\mathrm{BH}<\num{e8}\msun$ and have high luminosities $L\gtrsim0.1\,\ledd$ compared with BLS1s \citep[e.g.][] {Boller+96,Grupe&Mathur+04,2007ASPC..373..529Y,2017ApJS..229...39R,Waddell+20}, where $\ledd=1.26\times10^{38}(M_\mathrm{BH}/M_\odot)\,\lumcgs$ is the Eddington luminosity. $L/\ledd$ is likely to be interpreted as the eigenvector 1, a basic parameter that describes the relationship between the properties such as the X-ray spectral index $\alpha_\mathrm{X}$, the \ion{Fe}{ii} strength, and FWHM(H$\beta$) \citep{Boroson&Green+92,Boroson+02,Grupe+04}, and is sometimes interpreted as the ``age" of the AGN \citep{Grupe+99a}. In this context, NLS1s can be regarded as AGNs with a high $L/\ledd$ in the early phase of their evolution.

\subsection{X-ray View of NLS1s and Soft X-ray Excess}


The temperature of the accretion disk in an AGN is $\sim10^{5}\,\rm K$, which results in blackbody emission that peaks in the far ultraviolet (UV) band. However, due to strong Galactic extinction, a majority of the spectral energy distribution (SED) in this band cannot be observed. In the innermost region, a compact, hot ($kT\sim\ $100 keV), and optically thin corona is responsible for the hard X-ray power-law continuum due to inverse Comptonization of the disk photons \citep{1991ApJ...380L..51H,2015MNRAS.451.4375F}. The Comptonized photons also illuminate the disk where they are reprocessed within a Thompson optical depth and produce a reflection spectrum. This component includes a series of emission lines and a Compton hump peaking at $20-40\,\rm keV$. The most prominent feature in the reflection spectrum is the iron K$\alpha$ emission line around 6.4 keV \citep{Ross&Fabian+93,2002MNRAS.331L..35F,Reynolds&Nowak+03,jiang18}.

NLS1s show unique properties in X-ray bands such as steep soft X-ray spectra at $0.2-2\,\rm keV$ and extreme X-ray variability \citep{Boller+96,Leighly+99a,Grupe+01,2018rnls.confE..34G}. \citet{Gallo+06} defined two samples of NLS1s, the ``complex" sample, which shows significant spectral complexity and weaker X-ray luminosity at $2.5-10$ keV, and the ``simple" sample that does not strongly deviate from a power-law continuum. The ``complex" samples can+ either be explained by a partial-covering absorption model that obscures the X-ray emitting region \citep{2004PASJ...56L...9T}, or a light-bending model in the disk-reflection scenario \citep{2004MNRAS.349.1435M}.

The soft X-ray excess is a common phenomenon in many NLS1s and remains an active research topic. It refers to the excess of X-rays below 2 keV with respect to the extrapolation of the hard X-ray continuum. Studies of AGN samples have demonstrated that the soft excess is ubiquitous in both NLS1s and BLS1s, with stronger soft excess strength in NLS1s \citep[e.g. ][]{2007MNRAS.381.1426M,2009A&A...495..421B,gliozzi+2020}. 

There are now two popular competing models that are able to account for the soft excess. One is the warm Comptonization model. It assumes a warm ($kT_\mathrm{e}\sim\ 0.5-1$ keV) and optically thick ($\tau\sim\ 5-10$) corona in addition to the hot corona.
The Comptonization of the disk UV photons in the warm corona is responsible for the soft excess \citep[e.g.][]{2009MNRAS.398L..16J,2012MNRAS.420.1825J,Petrucci+18}. An alternative model is the relativistic blurred disk-reflection model. The emission lines in the soft X-ray band from the reflection component are blurred due to relativistic effects near the black hole and thus constitute the soft excess \citep[e.g.][]{Crummy+06,2020MNRAS.498.3888J}. The reflection model is also supported by the evidence of soft X-ray reverberation lags \citep{Fabian+09,Alston+20}.

In addition to explaining the soft excess, the relativistic reflection models can also provide a window to understand the innermost geometry of NLS1s and the physics of accretion \citep{Bambi+21,Reynolds+21}. Black hole spin ($a_*$) measurement is an active topic, and it is still an ongoing effort. One of the most credible and robust techniques for measuring the black hole spin is to model the relativistic reflection features. The reflection spectrum will be distorted and broadened by relativistic effects of Doppler broadening from the fast orbital motion, the gravitational redshift, and the light-bending due to the black hole's strong gravitational field \citep{1989MNRAS.238..729F,2002ApJ...570L..69M,2007ARA&A..45..441M}. There are various spin measurements of the supermassive black holes in AGNs, e.g. MCG$-$6$-$30$-$15 \citep[$a_*=0.91^{+0.06}_{-0.07}$; ][]{Marinucci+14}, Mrk 335 \citep[$a_*=0.83^{+0.10}_{-0.13}$; ][]{2014MNRAS.443.1723P}, IRAS 13224$-$3809 \citep[$a_*>0.99$; ][]{Jiang+18}, H1821$+$643 \citep[$a_*=0.62^{+0.22}_{-0.37}$; ][]{Sisk-Reynes+22} using the X-ray reflection technique. Despite that these measurements inevitably suffer from several systematic effects \citep{Laor+19}, the relativistic reflection is a robust signature. Apart from spin measurement, it is capable of probing the physical properties in the innermost region where strong relativistic effects dominate. As we observe the hard X-ray component produced by the hot compact corona, we can estimate the properties and the geometry of the corona \citep[e.g. ][]{2005ApJ...635.1203M,2012MNRAS.424.1284W,2018MNRAS.474.1538P}.

Apart from testing different physical models, an alternative approach to study the soft excess would be to quantitatively identify the soft excess strength regardless of the spectral components \citep[e.g.][]{2005A&A...432...15P,Boissay+16}. \citet{gliozzi+2020} (hereafter GW+20) conducted a model-independent correlation analysis of a clean sample of 68 objects (46 BLS1s and 22 NLS1s) in which they quantified the soft excess strength. However, this study only included NLS1s in the local universe (median $z=0.08$, with one outlier at $z=0.24$) using the $0.5-10$ keV spectra. We want to extend the study to higher redshifts.

In this study, we collect six NLS1s at redshift $0.35 < z < 0.92$ with sufficient count rates from the literature. We want to first examine whether these NLS1s at higher redshift have prominent soft excess emission. If they do, then we will further quantify the soft excess strength and compare the results with the ones in GW+20. In particular, the NLS1s at higher redshifts in our sample have relatively higher black hole masses than the NLS1 sample in GW+20. It is, therefore, interesting to compare the strength of the soft excess emission in these two samples. Although our NLS1s' observations do not have high enough signal-to-noise ratios (SNRs) to distinguish different soft excess modes as in previous work \citep[e.g., ][]{Middleton+07,Boissay+16,Jiang+18,Garcia+19}, we tentatively test for the warm corona and relativistic reflection models and find whether the two models provide consistent explanations to the soft excess emission of our sample as before, e.g., similar warm corona temperatures or similar disk densities.

\begin{figure}
    \centering
    \includegraphics[width=\columnwidth]{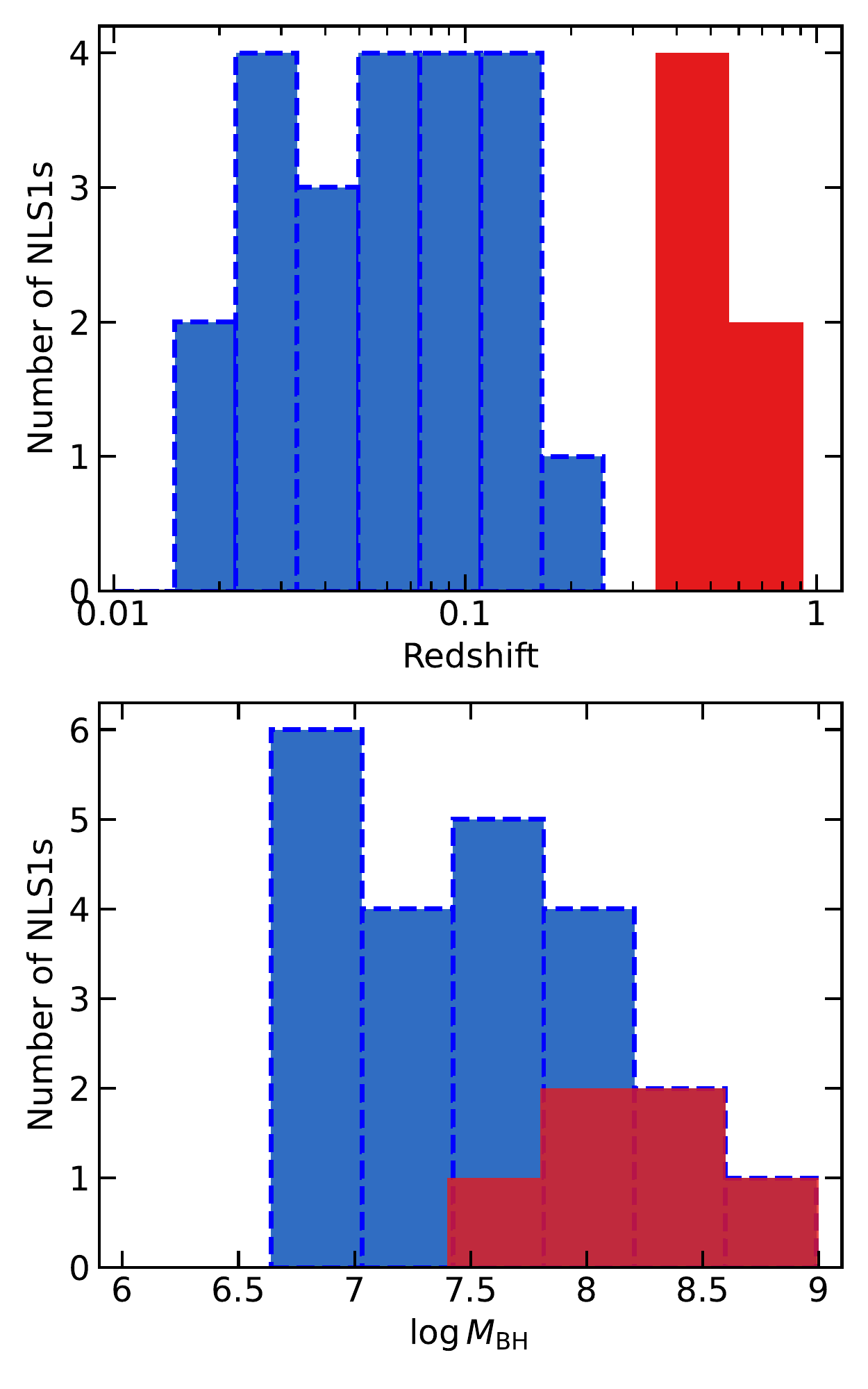}
    \caption{The distribution of the redshifts and the BH masses in our sample (in red). The NLS1s at lower redshifts in GW+20 are shown in blue with dashed edges.}\label{fig:hist}
\end{figure}

The paper is structured as follows. Section~\ref{sec:reduction} describes our sample and data reduction. In Section~\ref{sec:analysis}, we present our analysis and discussions. Specifically, in the first two subsections, we introduce our baseline models and illustrate the soft excess in our sample. In Section~\ref{subsec:sx_strength}, we quantify the soft excess and compare the results with the lower-redshift sample in GW+20. In Section~\ref{subsec:systematic}, we discuss the systematic uncertainties and the selection effects in our analysis. In Section~\ref{sec:discussion}, we apply the two possible physical models for the soft excess: the warm Comptonization model and the relativistic reflection model to see whether we can have further physical interpretations. Finally, in Section~\ref{sec:conclusion} we draw our conclusions and point out prospects.

\section{Observation Sample and Data Reduction}\label{sec:reduction}

\begin{table*}
\begin{threeparttable}
    \caption{List of NLS1s analyzed in this work.}
    \label{tab:src-list}
\begin{tabular}{llcccccc}
\hline
Source & Full Name & $z$ & $\log\mbh/\msun$  &  \xmm & Date & Net Exposure\tnote{\textit{e}} & Galactic $N_\mathrm{H}$ \\
& & & & ObsID & & ks & $\times10^{20}\,\mathrm{cm^{-2}}$\\
\hline
E1346+266 & E1346+266 & 0.92 & \num{8.63+-0.13}\tnote{\textit{a}} & 0109070201 & 2003-01-13 & 48, 55, 55 &  1.25 \\
PG 1543+489 & PG 1543$+$489 & 0.40 & \num{8.13+-0.33}\tnote{\textit{b}} & 0153220401 & 2003-02-08 & 9, 12, 12 &  1.67\\
& &  & & 0505050201 & 2007-06-09 & 11, 16, 17 &  \\
& & &  & 0505050701 & 2007-06-15 & 13, 16, 16 & \\
SDSS J020435 & SDSS J020435.18$-$093154.9 & 0.62 & \num{8.20+-0.31}\tnote{\textit{c}} & 0763910301 & 2015-08-08 & $-$, 11, 12 & 2.42 \\
SDSS J024651 & SDSS J024651.91$-$005930.9 & 0.47 & \num{8.29+-0.15}\tnote{\textit{c}} & 0402320101 & 2007-02-02 & 8, 11, 11 & 4.27\\
WISEA J033429 & WISEA J033429.44$+$000610.9 & 0.35 & \num{7.83+-0.13}\tnote{\textit{d}} & 0402320201 & 2007-02-02 & 8, 10, 10& 9.85 \\
1WGA J2223 & 1WGA J2223.7$-$0206 & 0.46 & \num{7.55+-0.14}\tnote{\textit{d}} & 0090050601 & 2001-12-06 & 15, $-$, $-$ & 5.70 \\
\hline
\end{tabular}
    \begin{tablenotes}
    \item Notes: The BH masses are measured based on the line width of $^\mathrm{\textit{a}}$ \ion{Mg}{ii} \citet{2011ApJS..194...45S}; $^\mathrm{\textit{b}}$ H$\mathrm{\beta}$ \citet{2008ApJS..176..355K}; $^\mathrm{\textit{c}}$ \ion{Mg}{ii} \citet{2009ApJ...707.1334W}. Based on $^\mathrm{\textit{d}}$ RM method with H$\beta$, \citet{2000ApJ...533..631K,2004A&A...418..907G}. $^\mathrm{\textit{e}}$ The net exposure is reported in the order of EPIC-pn, EPIC-MOS1, and EPIC-MOS2. SDSS~J020435 lacks pn data; 1WGA~J2223 lacks MOS1 and MOS2 data.
    \end{tablenotes}
\end{threeparttable}
\end{table*}

We select six NLS1s with redshift $0.35<z<0.92$ observed by the European Space Agency's \xmm\ satellite \citep{2001A&A...365L...1J}. Details about the observations are summarized in Table \ref{tab:src-list}. Their redshift and BH mass distributions are shown in Figure~\ref{fig:hist}. The studied redshift range is extended much beyond the one in the previous sample. The BH masses are distributed at the upper end of the previous sample, which are estimated by the Reverberation Mapping (RM) method or the emission-line properties (see the notes in the Table for details). In addition, we include the FWHM(H$\beta$) and \ion{Fe}{ii}/H$\beta$ ratios from the literature in Table~\ref{tab:optical-uv-xray}. We also show the full list of our candidate samples with their EPIC-pn count rates in Appendix~\ref{app:candidates}. We do not aim to conduct a thorough study of the entire population of NLS1s at the given redshift range. Instead, we want to examine how these selected NLS1s behave given their higher redshift and the $M_\mathrm{BH}$ at the upper end of the local sample. Future systematic studies of the NLS1s at higher redshift can be carried out with archived \xmm\ data of NLS1 in catalogs \citep[e.g., ][]{Rakshit+17}

We use the European Photon Imaging Camera (EPIC) observations collected by EPIC-pn and EPIC-MOS \citep{2001A&A...365L..18S,2001A&A...365L..27T}. Note that SDSS~J020435 only has available EPIC-MOS data; 1WGA~J2223 only has available EPIC-pn data. Both sources fall into the gaps of the CCD chips in the FoV of either EPIC-pn or EPIC-MOS. We use the \textit{XMM-Newton} Science Analysis System (SAS) v20.0.0 software package to reduce the data. We first generate the clean event files using \texttt{epproc} (\texttt{emproc}) for EPIC-pn observations (EPIC-MOS observations). We then extract the good time intervals (GTIs) by selecting the time intervals with weak flaring particle backgrounds. For PG~1543+489, due to strong background flaring in all observations, we apply the iterative 3$\sigma$ cleaning procedure for the EPIC $10-15$ keV range described in \citet{vignali+08}. For other observations, we use the standard EPIC data reduction thread, where we select the single event (PATTERN=0) count rate in the $\geq10$ keV band that is smaller than 0.40 $\rm counts\,s^{-1}$ (0.35 $\rm counts\,s^{-1}$) for EPIC-pn (EPIC-MOS) data. Then we filter the event lists for all observations by selecting single and double events for EPIC-pn (PATTERN$\le$4) and EPIC-MOS (PATTERN$\le$12) in a circular region. No pile-up issue is found in our observations. The backgrounds are extracted in nearby regions on the same chip. The redistribution matrix files and ancillary response files are generated by \texttt{rmfgen} and \texttt{arfgen}.

We use EPIC spectra in the $0.3-10$ keV energy range in our analysis. Given the redshift range of our sample, the analyzed rest-frame energy band is consistent with the GW+20 local samples which were analyzed in the $0.5-10$ keV band. The lightcurves of our samples do not show strong X-ray variability. For PG~1543+489, we use the \texttt{epicspeccombine} tool to stack the three observations from EPIC-pn, EPIC-MOS1, and EPIC-MOS2 spectra, respectively. We group the spectra using the \texttt{specgroup} command so that each bin has a minimum signal-to-noise ratio (SNR) of 3 (minimum SNR of 6 for PG~1543+489) and a minimum width of $1/3$ of the full width half maximum resolution at the central energy of the group.

\section{Spectral Analysis}\label{sec:analysis}

We use \texttt{Xspec} v12.12.0 \citep{1996ASPC..101...17A} and the $\chi^2$ statistics in our analysis. The Galactic hydrogen column density $N_\mathrm{H}$ is calculated according to \citet{2013MNRAS.431..394W}, which is shown in Table~\ref{tab:src-list}.
The luminosity distance $d_\mathrm{L}$ is retrieved from the NED website\footnote{https://ned.ipac.caltech.edu/}, where $H_0=67.8\,\rm km\,s^{-1}\,Mpc^{-1}$, $\Omega_\mathrm{matter}=0.308$, and $\Omega_\mathrm{vacuum}=0.692$ are assumed.

\subsection{Soft Excess Emission}

\begin{figure*}
    \centering
    \includegraphics[width=\textwidth]{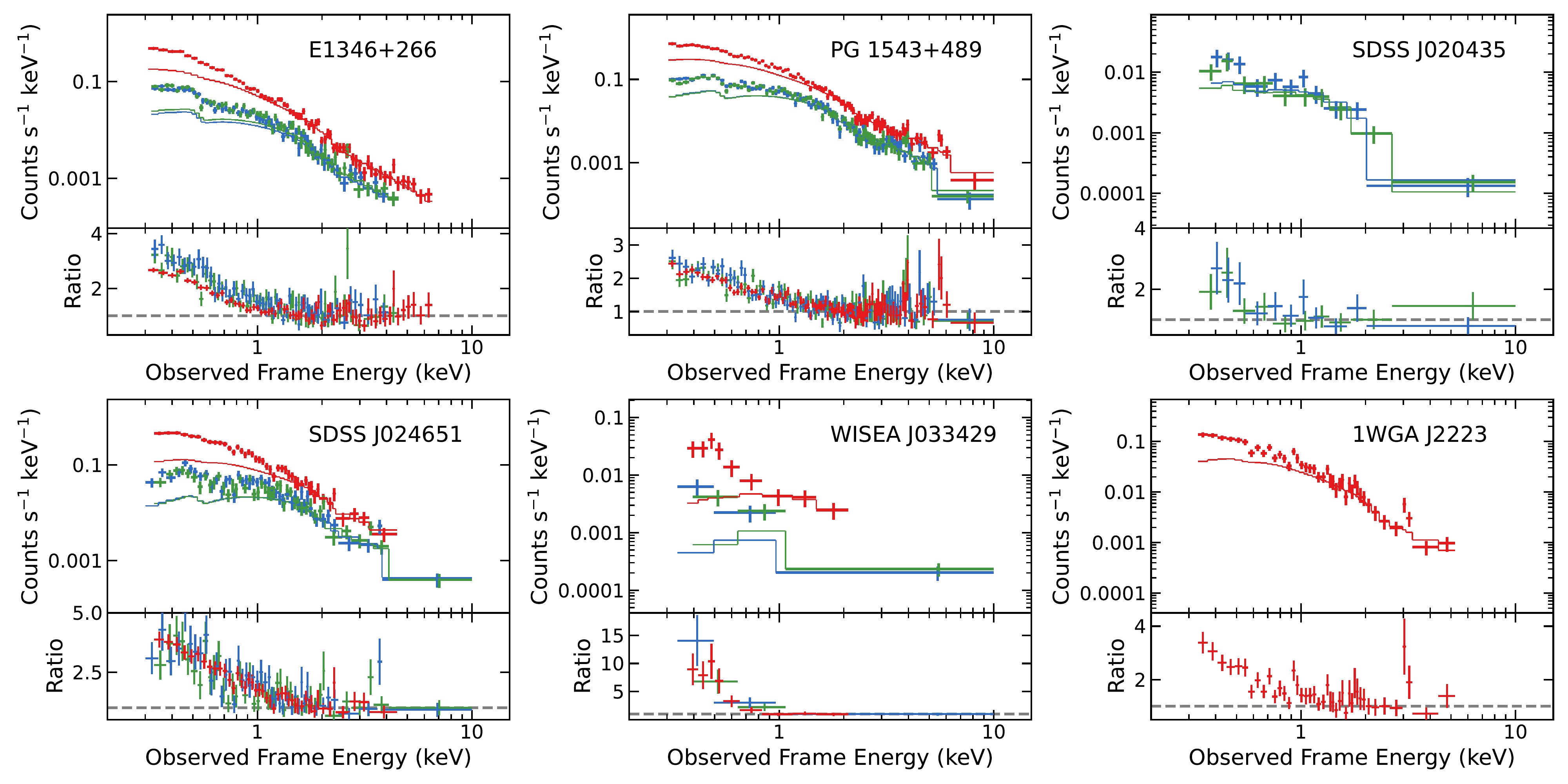}
    \caption{Count spectra and data-to-model ratio plots showing the soft excess of the NLS1s in this work. The upper plots show the count spectra and the corresponding absorbed powerlaw models. The lower plots show the data-to-model ratio. The red, blue, and green data (lines) correspond to pn, MOS1, and MOS2 data (models).}
    \label{fig:sx_strength}
\end{figure*}

We fit the $2-10$ keV energy band in the observed frame with an absorbed powerlaw at the Galactic column density values ($0.7-10$ keV for SDSS~J020435 and WISEA~J033429 due to inadequate data bins in the $2-10$ keV band). We then include the $0.3-2$ keV band data \textit{without} re-fitting to illustrate the soft excess emission in our sources. We show the data/model ratio plots in Figure~\ref{fig:sx_strength}. All spectra show strong excess emission in the soft X-ray band below 1 keV. To model the soft excess component, we continue to apply a baseline model to the spectra.

\subsection{Baseline Models and Eddington Ratio}\label{subsec:baseline}

We apply a phenomenological baseline model across the available energy band of $0.3-10$~keV in the observed frame. Our baseline model is
\begin{equation*}
    \texttt{const$\times$TBabs$\times$zashift$\times$(bbody+powerlaw)}
\end{equation*}
in \texttt{Xspec} notation. 
We use \texttt{powerlaw} to model the high-energy Comptonized continuum, and we use the blackbody model \texttt{bbody} with the temperature at $kT$ to fit the soft excess emission. The \texttt{const} is to account for the flux calibration uncertainties. The \texttt{zashift} model accounts for the redshift. The \texttt{Tbabs} model describes the absorption along the line-of-sight. As a first attempt, we allow the column density $N_\mathrm{H}$ to vary. We shall show all our spectral plots in the rest frame of the sources. 

The iron K$\alpha$ emission feature around 6.4 keV is produced by the illumination of the disk by the primary X-ray source and the fluorescence of Fe K-shell electrons \citep{1989MNRAS.238..729F,2007ARA&A..45..441M}. This feature is not prominent in our sample due to the low signal-to-noise ratio (SNR) in the iron emission band. Thus, we cannot conclude the existence of the iron K$\alpha$ line in most objects. However, \citet{vignali+08} found evidence of a broad iron K$\alpha$ emission line in PG1543$+$489, so we consider an additional component for the iron K$\alpha$ emission for this source in our analysis by following the same approach as in \citet{vignali+08}.

For PG~1543+489, we add a \texttt{relline} \citep{2010MNRAS.409.1534D} component to model the relativistic iron K$\alpha$ emission feature. We set the rest-frame emission line energy $E_\mathrm{line}$ at $6.7\,\rm keV$ as it is the best-fit value obtained by \citet{vignali+08}. The illumination pattern onto the disk, namely the emissivity profile $\epsilon(r)$ is assumed to be a broken-powerlaw: $\epsilon(r)\propto r^{-q_\mathrm{in}}$ for $r<R_\mathrm{br}$, and $\epsilon(r)\propto r^{-3}$ for $r>R_\mathrm{br}$, where $q_\mathrm{in}$ is the inner emissivity index and $R_\mathrm{br}$ is the breaking radius. We also allow a free BH spin $a_*$ and inclination angle $\theta$.

\begin{figure*}
    \centering
    \includegraphics[width=\textwidth]{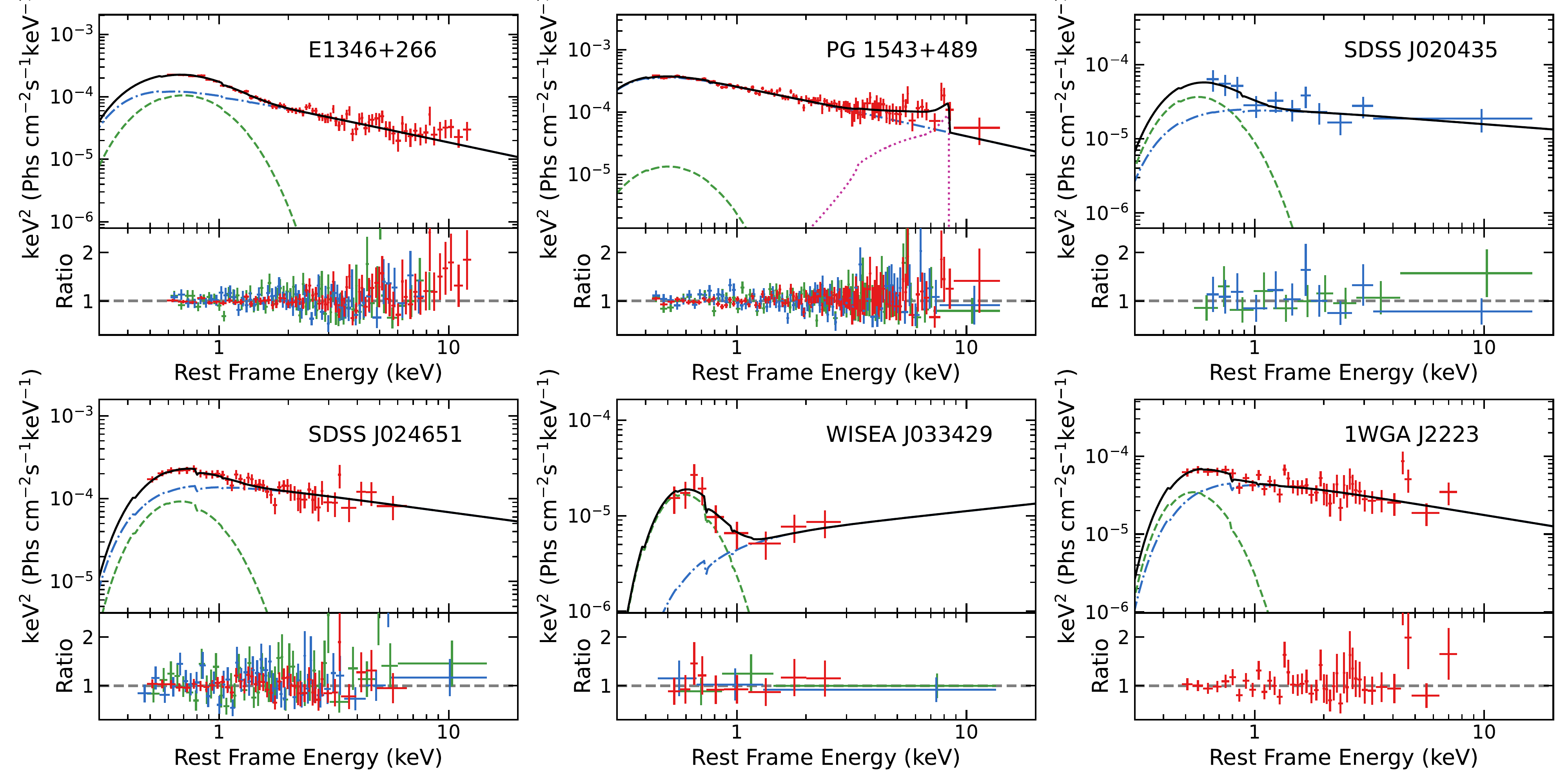}
    \caption{Baseline model fits. Unfolded best-fit model plots and data-to-model ratio plots for each NLS1. The \texttt{bbody} model is used to fit their soft excess emission. The upper plots show the best-fit models. We only show the pn data in red for clarity (MOS1 data in blue for SDSS J020435). Solid black lines: the total model; blue dash-dotted lines: \texttt{powerlaw}; green dashed lines: \texttt{bbody}. The lower plots show the data-to-model ratios. The red, blue, and green data points correspond to pn, MOS1, and MOS2 data.}
    \label{fig:baseline}
\end{figure*}

Figure~\ref{fig:baseline} shows the best-fit models and data/model ratio plots for the six sources. Table~\ref{tab:baseline} gives the best-fit results using the baseline models. We find that the $N_\mathrm{H}$ cannot be constrained in our fits, so we fix the $N_\mathrm{H}$ to the Galactic values throughout our analysis. We also report the $N_\mathrm{H}$ upper limit within 90\% confidence level in column (13) of Table~\ref{tab:optical-uv-xray}. None of the sources are highly obscured. Based on the best fits, we measure the X-ray luminosity $L_\mathrm{X}$ from $2-10\,\rm keV$ in the rest frame after correcting for the Galactic absorption. We calculate the logarithmic Eddington ratio $\log\leddrat=\log(L_\mathrm{bol}/L_\mathrm{Edd})$, where $L_\mathrm{bol}$ is the bolometric luminosity, with the BH mass given in Table~\ref{tab:src-list}. The bolometric correction factor \kbol\ linking the $2-10\,\rm keV$ $L_\mathrm{X}$ and the $L_\mathrm{bol}$ is estimated by the approximation in \citet{netzer+19}:
\begin{equation}
    k_\mathrm{bol}=7\times(L_\mathrm{X,2-10\,keV}/10^{42}\lumcgs)^{0.3}.\label{eqn:kbol}
\end{equation}
We also present the measurements of $\log\leddrat$ by optical observations in the literature. The results are summarized in columns (7), (8), and (9) of Table~\ref{tab:optical-uv-xray} (see the notes therein for more information). All sources present large $\log\leddrat>-1$ estimated from the X-rays except for WISEA~J033429. The large Eddington ratio of our sample is consistent with the typical NLS1 behavior that they are actively accreting matter.

\begin{table*}
\begin{threeparttable}
\caption{The optical, UV, and X-ray properties of our NLS1 sample. (1) Source Name; (2) \ion{Fe}{ii}/H$_\mathrm{\beta}$; (4) rest-frame \SI{2500}{\angstrom} flux density ($\times10^{-27}\,\fluxdenscgs$); (5) rest-frame \SI{2}{keV} flux density ($\times10^{-31}\,\fluxdenscgs$); (6) $\alpha_\mathrm{OX}$; (7) rest-frame $2-10\ \si{keV}$ X-ray luminosity ($\times10^{43}\,\lumcgs$); (8) X-ray Eddington ratio measured by Equation~\ref{eqn:kbol}; (9) Optical Eddington ratio reported in literature; (10) \sx{1}; (11) \sx{2}; (12) \sx{3}; (13) $N_\mathrm{H}$ upper limit ($\times10^{20}\,\rm cm^{-2}$).}
    \label{tab:optical-uv-xray}
\begin{tabular}{lcccccccclllc}
\hline
(1) & (2) & (3) & (4) & (5) &  (6) & (7) & (8) & (9) & (10) & (11) & (12) & \textbf{(13)}\\
Source  & \ion{Fe}{ii}/H$\beta$ & FWHM(H$\beta$) & $f_\mathrm{UV}$ & $f_\mathrm{X}$ & $\alpha_\mathrm{OX}$ & $L_\mathrm{X}$ & $\log\lambda_\mathrm{Edd,X}$ & $\log\lambda_\mathrm{Edd,opt}$  & \sx{1} & \sx{2}  & $\sx{3}$  &  $N_\mathrm{H}$ \\
\hline
E1346+266 & 0.98 & 1840 & 0.30 & 2.16 & -1.21 & 42.5 & $-0.465$ & $-0.687$\tnote{\textit{a}} & $0.62_{-0.12}^{+0.11}$ & $0.44_{-0.08}^{+0.07}$ & $13.5_{-2.2}^{+1.8}$ & 1.6 \\
PG 1543+489 & 0.86 & 1560 & 10.0 & 5.03 & -1.65 & 16.6 & $-0.495$ & $0.369$\tnote{\textit{b}} & $0.017_{-0.017}^{+0.005}$ & $0.02_{+0.03}^{+0.01}$ & $0.4_{-0.5}^{+0.1}$  & 2.9\\
SDSS J020435 & 0.83 & 1905 & 1.40 & 0.74 & -1.64 & 8.5 & $-0.94$ & $-0.466$\tnote{\textit{c}} & $0.7_{-0.7}^{+0.5}$ & $0.4_{-0.3}^{+0.2}$ & $4_{-4}^{+2}$  &  29 \\
SDSS J024651 & 0.49 & 1725 & 3.83 & 4.04 & -1.53 & 21 & $-0.52$ & $-0.356$\tnote{\textit{c}} &  $0.41_{-0.24}^{+0.15}$ & $0.24_{-0.14}^{+0.08}$ & $5_{-3}^{+2}$  & 7.9 \\
WISEA J033429 & 0.69 & 2127 & 3.46 & 0.23 & -1.99 & 1.4 & $-1.60$ & $-0.551$\tnote{\textit{c}} & $2.1_{-1.9}^{+1.3}$ & $0.8_{-0.5}^{+0.4}$ & $2.0_{-1.0}^{+0.6}$  &  91 \\
1WGA J2223 & --- & 2000 & 1.09 & 1.21 & -1.52 & 8.3 & $-0.31$ & $>-0.170$\tnote{\textit{d}} & $0.36_{-0.20}^{+0.14}$ & $0.23_{-0.12}^{+0.08}$ &  $8_{-4}^{+3}$  &  13 \\
\hline
\end{tabular}
\begin{tablenotes}
    \item Notes: The optical Eddington ratios are collected from $^\mathrm{\textit{a}}$ \citet{2011ApJS..194...45S}. $^\mathrm{\textit{b}}$ \citet{2004MNRAS.350L..31B}. $^\mathrm{\textit{c}}$ \citet{Rakshit+17}. $^\mathrm{\textit{d}}$ \citet{2004A&A...418..907G}.
\end{tablenotes}
\end{threeparttable}
\end{table*}

\subsection{The Soft Excess Strength: Comparison with BLS1s and NLS1s at lower redshifts}\label{subsec:sx_strength}

Literature shows a number of ways to quantify the soft excess strength \citep[e.g.][]{2005A&A...432...15P,2013A&A...549A..73P,Boissay+16}. Here we adapt the three quantities defined in GW+20 to characterize the soft excess strength.

\sx{1} is defined as the ratio of the unabsorbed $0.5-2$ keV flux in the rest frame of the blackbody component over the Comptonized component:

\begin{equation*}
    \sx{1}=\qty(\frac{F_\mathrm{bb}}{F_\mathrm{cp}})_\mathrm{0.5-2\ keV}
\end{equation*}

Similarly, \sx{2} is the ratio of the unabsorbed rest-frame $0.5-10$ keV flux of the blackbody component over the Comptonized component:

\begin{equation*}
    \sx{2}=\qty(\frac{F_\mathrm{bb}}{F_\mathrm{cp}})_\mathrm{0.5-10\ keV}
\end{equation*}

Finally, \sx{3} is the ratio of the unabsorbed rest-frame $0.5-2$ keV blackbody luminosity over the Eddington luminosity:

\begin{equation*}
    \sx{3}=\frac{L_\mathrm{bb,0.5-2\ keV}}{\ledd}
\end{equation*}

We measure \sx{1}, \sx{2}, and \sx{3} using the baseline models defined in Section~\ref{subsec:baseline}, where \texttt{bbody} is the phenomenological description of the soft excess, and the Comptonized component is modeled by \texttt{powerlaw}. All flux measurements are based on the model extrapolation in the rest-frame energy band. Columns (10), (11), and (12) in Table~\ref{tab:optical-uv-xray} summarize the measurements of the soft excess strength.

\begin{table*}
\footnotesize
\caption{Best-fit table of the baseline models. The flux $F$ is in units \fluxcgs. The reported uncertainties correspond to the 90\% confidence level for one parameter ($\Delta\chi^2=2.71$). * indicates that the parameter is frozen in the fit.}
\label{tab:baseline}
\begin{tabular}{lcccccc}
\hline
\hline
& E1346+266 & PG1543+489 & SDSS J0204 & SDSS J0246 & WISEA J033 & 1WGA J2223\\
\hline
\texttt{bbody} &  &  & & & & \\
$kT\,\mathrm{(keV)}$ & $0.153_{-0.009}^{+0.008}$ & $<0.16$ & $0.11_{-0.06}^{+0.07}$ & $0.14_{-0.02}^{+0.02}$ & $0.09_{-0.03}^{+0.03}$ & $0.08_{-0.03}^{+0.03}$\\
$\log F^\mathrm{bb}_\mathrm{0.5-10\,keV}$ & $-12.80_{-0.07}^{+0.06}$ & $-13.8_{-0.7}^{+0.2} $ & $-13.4_{-0.5}^{+0.2}$ & $-12.83_{-0.24}^{+0.14}$ & $-13.40_{-0.20}^{+0.13}$ & $-13.37_{-0.23}^{+0.15}$\\
\hline
\texttt{powerlaw} &  &  & & & & \\
$\Gamma$ & $2.78_{-0.09}^{+0.09}$ & $2.80_{-0.08}^{+0.04}$ & $2.2_{-0.4}^{+0.4}$ & $2.37_{-0.18}^{+0.18}$ & $1.7_{-0.8}^{+0.8}$ & $2.48_{-0.21}^{+0.19}$\\
$\log F^\mathrm{pl}_\mathrm{0.5-10\,keV}$ & $-12.44_{-0.03}^{+0.03}$ & $-12.08_{-0.01}^{+0.01}$ & $-12.96_{-0.12}^{+0.09}$ & $-12.22_{-0.03}^{+0.03}$ & $-13.30_{-0.14}^{+0.19}$ & $-12.72_{-0.04}^{+0.03}$\\
\hline
\texttt{relline} &  &  & & & & \\
$E_\mathrm{line}$ (keV) &  & $6.7^{*}$ & & & & \\
$q_\mathrm{in}$ &  & $>6$ & & & & \\
$q_\mathrm{out}$ &  & $3^*$ & & & & \\
$R_\mathrm{br}$ ($r_\mathrm{ISCO}$) &  & $2.1_{-0.3}^{+0.7}$ & & & & \\
$a_\mathrm{*}$ &  & $>0.97$ & & & & \\
Incl (deg) &  & $64_{-12}^{+11}$ & & & &\\
$\log F^\mathrm{relline}_\mathrm{0.5-10\,keV}$ & & $-13.0_{-0.2}^{+0.3}$ & & & & \\
\hline
$\chi^2/d.o.f.$ & 168.83/179 & 253.05/238 & 8.15/16 & 116.96/126 & 3.15/9 & 38.01/39\\
\hline
\hline
\end{tabular}
\end{table*}

\begin{figure}
    \centering
    \includegraphics[width=\columnwidth]{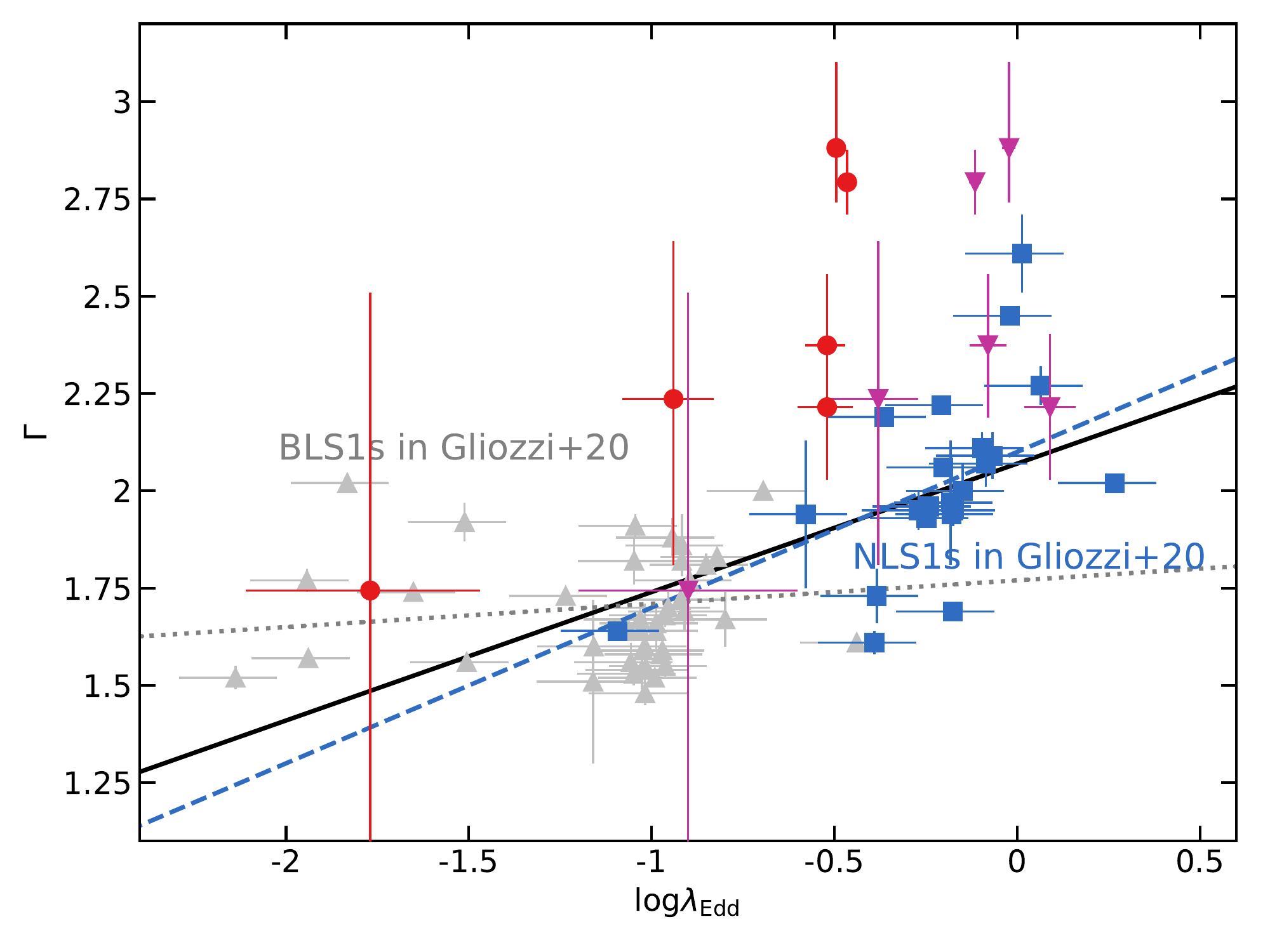}
    \caption{$\Gamma$ versus $\log\lambda_\mathrm{Edd}$. Red circles are our  NLS1 sample using the \kbol\ measured by Equation~\ref{eqn:kbol}; blue squares and grey up-pointing triangles are NLS1s and BLS1s at lower redshifts in GW+20; purple down-pointing triangles are our NLS1 sample using the \kbol\ in GW+20. The solid line represents the linear regression obtained for the clean NLS1s and BLS1s in GW+20.}
    \label{fig:gamma-mdot}
\end{figure}

\begin{figure*}
    \centering
    \includegraphics[width=\textwidth]{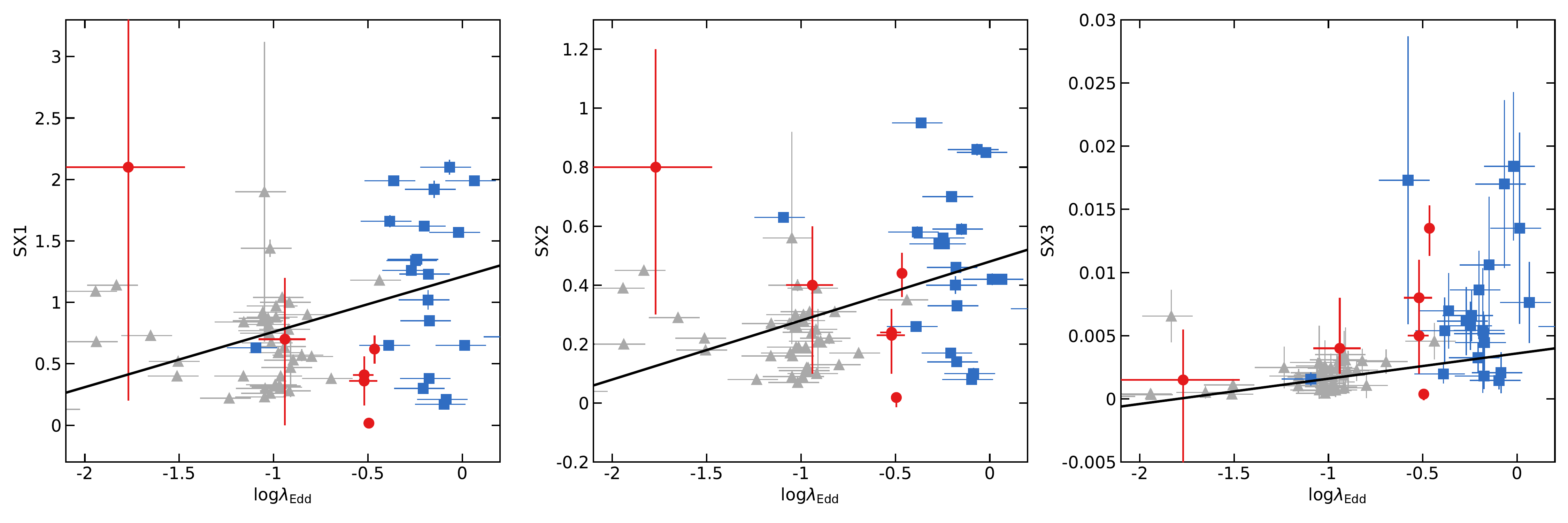}
    \caption{Soft excess strength versus $\log\lambda_\mathrm{Edd}$. Red circles are our high-redshift NLS1 sample; blue squares and grey triangles are NLS1s and BLS1s in GW+20. The solid line represents the linear regression obtained for the clean NLS1s and BLS1s in GW+20.}
    \label{fig:sx-mdot}
\end{figure*}

\begin{figure*}
    \centering
    \includegraphics[width=\textwidth]{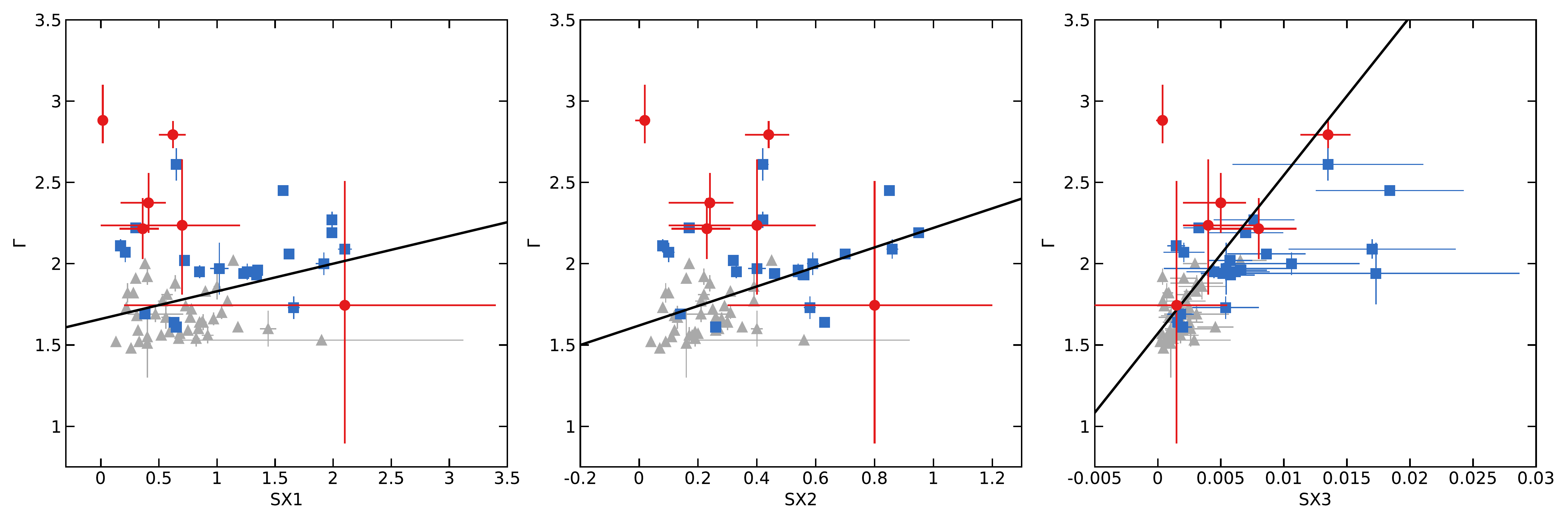}
    \caption{$\Gamma$ versus \sx{1}, \sx{2}, and \sx{3}. Red circles are our high-redshift NLS1 sample; blue squares and grey triangles are NLS1s and BLS1s in GW+20. The solid line represents the linear regression obtained for the clean NLS1s and BLS1s in GW+20.}
    \label{fig:gamma-sx}
\end{figure*}

We compare the results of our sample with the NLS1s and BLS1s at lower redshift in GW+20. While their NLS1s have a median $z=0.08$ with one outlier at $z=0.24$, our sample spans from $z=0.35$ to $z=0.92$, which is beyond the previous sample's range.  In Figures~\ref{fig:gamma-mdot}, \ref{fig:sx-mdot}, and \ref{fig:gamma-sx}, the red circles are our NLS1 sample at higher redshift; the blue squares and the grey triangles are the NLS1s and BLS1s measured in GW+20. Figure~\ref{fig:gamma-mdot} shows the correlations between the photon index $\Gamma$ and the $\log\leddrat$. Our NLS1s roughly follow the positive trend between $\Gamma$ and $\log\leddrat$ as revealed by the GW+20 sample. They also show similar photon indices to the lower-redshift sources, but the \leddrat\ is slightly lower. This comparison is band-consistent. Considering the difference in redshift, we conclude the spectra of our higher-redshift sample and the local sample in GW+20 have comparable energy ranges in the rest frame.

Figure~\ref{fig:sx-mdot} shows the soft excess strength versus the Eddington ratio. While the NLS1s in GW+20 show a variety of soft excess strengths, they present similar values of the Eddington ratio. This is also the case for the BLS1s in grey points. In particular, the NLS1s show relatively stronger soft excess strength than the BLS1s. Our NLS1s show a similar range of soft excess strength despite the relatively lower \leddrat. Among them, WISEA~J033429 presents the highest \sx{1} and \sx{2} and the second lowest \sx{3}. \sx{3} is imprinted with the absolute strength of the soft excess, as it compares the soft X-ray luminosity directly with the Eddington luminosity and thus is susceptible to the Eddington ratio. \sx{1} and \sx{2} can be a better indicator of the strength of the soft excess relative to the powerlaw continuum, regardless of the bolometric luminosity of the source. 

Figure~\ref{fig:gamma-sx} shows the photon index $\Gamma$ versus the soft excess strength. GW+20 presented a positive $\Gamma-\textit{SX}$ correlation in their AGN sample. Though our sources roughly follow a similar trend, we cannot be conclusive about a similar correlation due to the small number of sources. PG~1543+489 is the outlier in our sample. The soft excess strength is weak (lowest \sx{1}, \sx{2}, and \sx{3} in the sample), while it shows a very soft powerlaw continuum. It is also one of the most massive and luminous sources in our sample and the NLS1s in GW+20.

\begin{figure}
    \centering
    \includegraphics[width=\columnwidth]{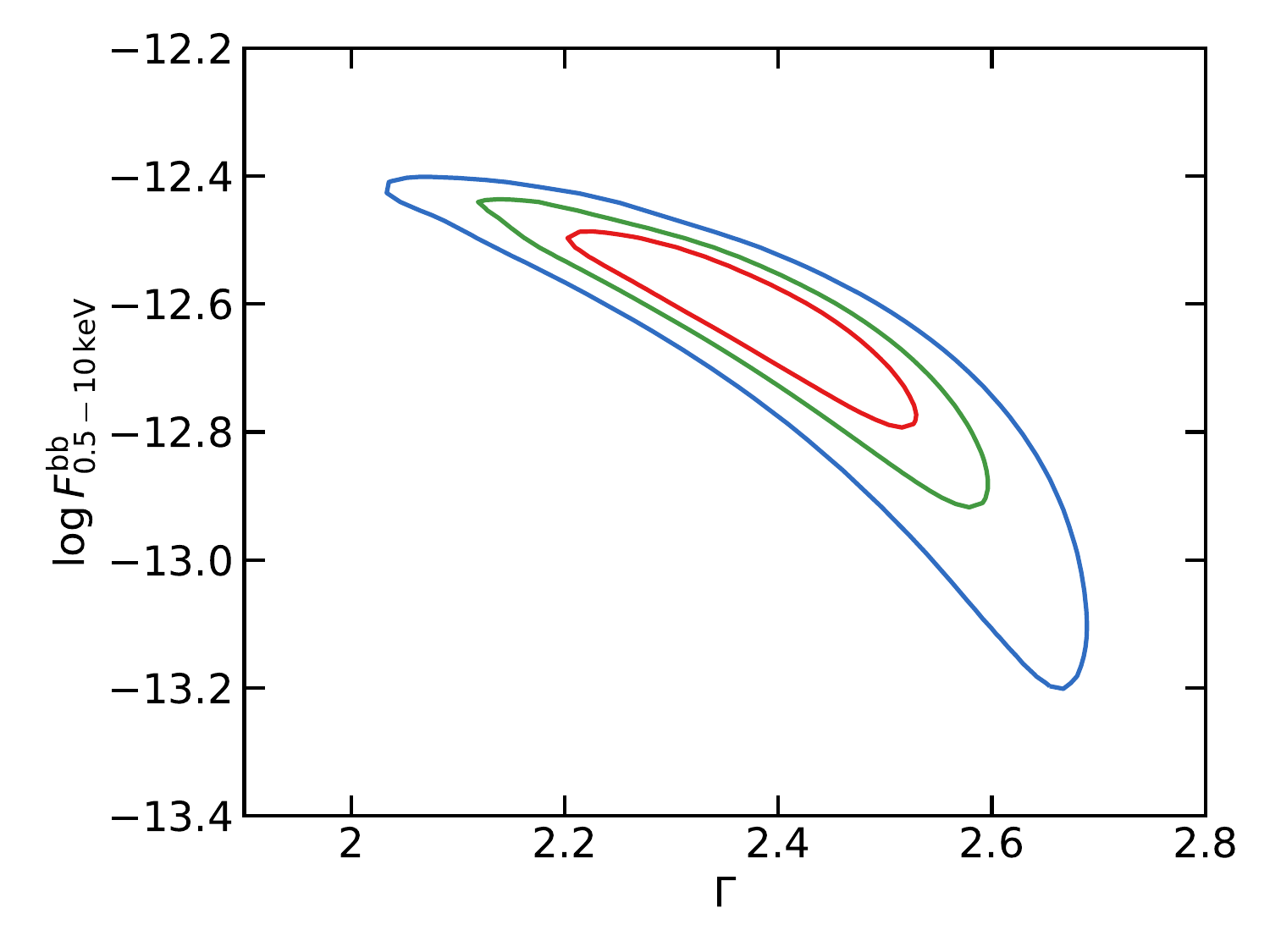}
    \caption{Constraints on the flux of the \texttt{bbody} component ($\log F^\mathrm{bb}_\mathrm{0.5-10\,keV}$) and the powerlaw photon index ($\Gamma$) for the source SDSS~J024651. The red, green, and blue curves represent 68\%, 90\%, and 99\% confidence levels for two relevant parameters, respectively. The banana-shaped contours clearly show a degeneracy between \texttt{bbody} and \texttt{powerlaw} component.}
    \label{fig:e1346_cont}
\end{figure}

The high redshifts of our sample move a great portion of the soft excess emission beyond the lower limits of the detection range. The lack of observation above 10 keV in the observed frame also makes it hard to accurately identify the continuum. These factors would likely cause a degeneracy between the soft excess component and the powerlaw component, and thus a large uncertainty in \sx{1} and \sx{2}. To better illustrate the degeneracy, we plot the constraints on the flux of the \texttt{bbody} component from $0.5-10$ keV and the powerlaw photon index for SDSS~J024651 in Figure \ref{fig:e1346_cont}. Although the confidence levels shown with the contours are consistent with the uncertainties in Table~\ref{tab:baseline}, the ``banana-shaped" contours clearly show the degeneracy between the soft excess flux and the photon index. Within 99\% confidence level, two possible solutions are found to explain the data with similarly good fits : (1) $\Gamma=2.6$ and $\log F^\mathrm{bb}_\mathrm{0.5-10\,keV}=-13.2$ (2) $\Gamma=2.0$ and $\log F^\mathrm{bb}_\mathrm{0.5-10\,keV}=-12.4$.

PG1543+489 is the only source in our sample that shows a blurred iron line feature. By modeling the emission line with the relativistic line emission model \texttt{relline}, we would be able to probe the innermost region of the accretion disk. The results indicate a fast-spinning BH ($a_*>0.97$). The emissivity profile shows a steep powerlaw ($q_\mathrm{in}>6$) within a relatively small radius ($R_\mathrm{br}=2.1_{-0.3}^{+0.7}$). This scenario indicates a strong illumination from the primary source onto the innermost disk region, where the relativistic effect can play a significant role in the outgoing reflection spectrum. This is consistent with the blurred iron line feature observed in this source.

\subsection{The $\alpha_\mathrm{OX}-L_\mathrm{\SI{2500}{\angstrom}}$ relation }\label{subsec:UV}

Apart from the X-ray observations, we also incorporate their historical observations at the optical and UV wavelengths. In particular, we focus on their rest-frame \SI{2500}{\angstrom} monochromatic luminosities $L_{\SI{2500}{\angstrom}}$ and optical-to-X-ray spectral indices $\alpha_\mathrm{OX}$. 

The $\alpha_\mathrm{OX}$ is defined as $\alpha_\mathrm{OX}\equiv\log(f_\mathrm{X} / f_\mathrm{UV})/\log(\nu_\mathrm{X}/\nu_\mathrm{UV})=0.3838\log(f_\mathrm{X} / f_\mathrm{UV})$, where $f_\mathrm{X}$ and $f_\mathrm{UV}$ are the \SI{2}{keV} and \SI{2500}{\angstrom} rest-frame flux density; $\nu_\mathrm{X}$ and $\nu_\mathrm{UV}$ are their corresponding frequencies. Previous studies have revealed a strong anti-correlation between $\alpha_\mathrm{OX}$ and $L_\mathrm{\SI{2500}{\angstrom}}$ \citep[e.g.,][]{Strateva+05,Just+07}. The ``simple" NLS1s defined by \citet{Gallo+06} follow the correlation relatively well and appear to be in a typical flux state, while the ``complex" NLS1 sample is representative of objects in X-ray low-flux states. They can be greatly below the nominal $\alpha_\mathrm{OX}-L_\mathrm{\SI{2500}{\angstrom}}$ relation and present stronger X-ray variability. 

\begin{figure}
    \centering
    \includegraphics[width=\columnwidth]{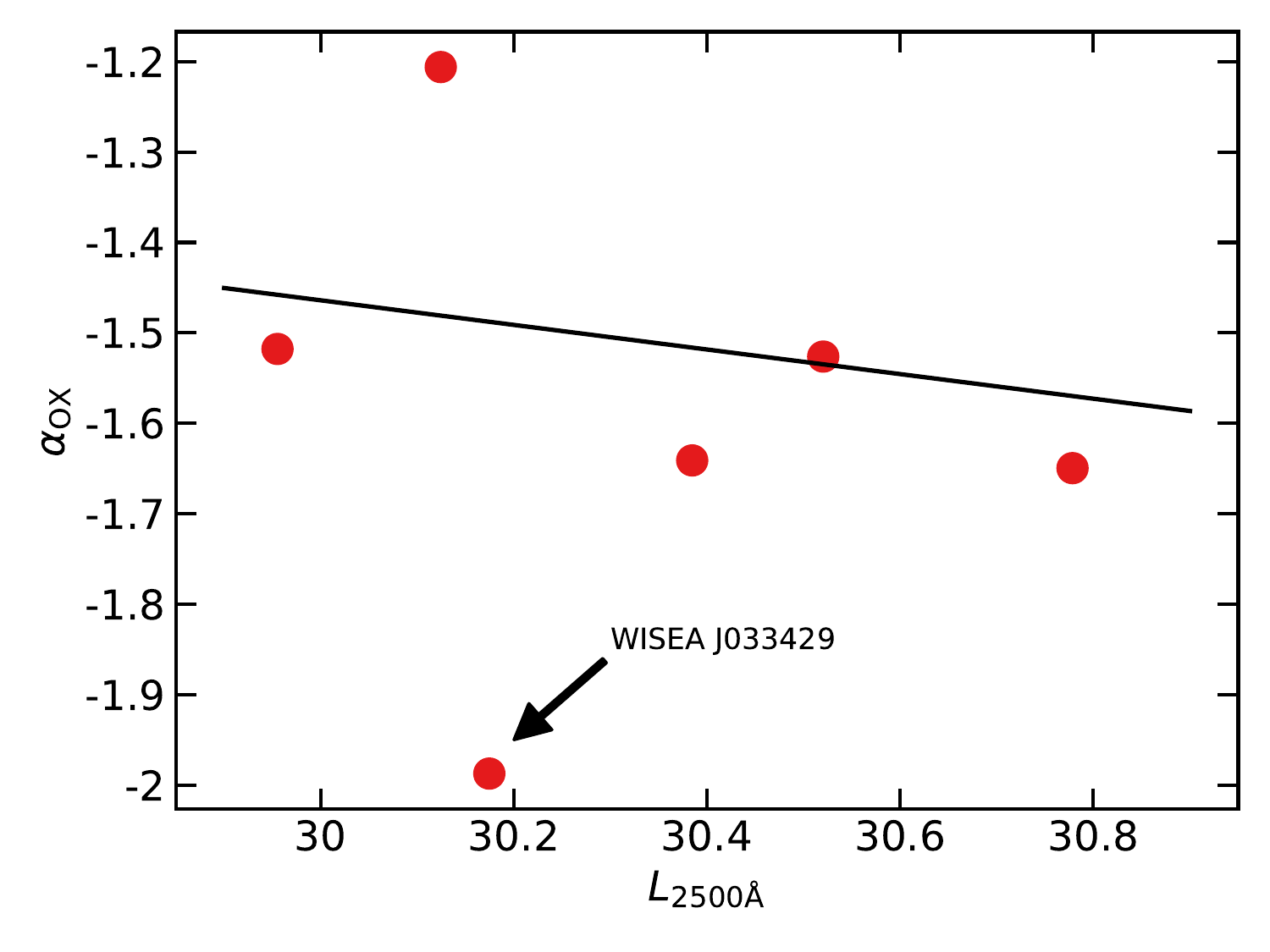}
    \caption{The $\alpha_\mathrm{OX}$ versus the \SI{2500}{\angstrom} monochromatic luminosity $L_\mathrm{\SI{2500}{\angstrom}}$. The red-filled circles are our NLS1s. The solid black line is the dependence of $\alpha_\mathrm{OX}$ for radio-quiet type-1 AGNs ($\alpha_\mathrm{OX}=-0.136L_\mathrm{\SI{2500}{\angstrom}}+2.616$) as in \citet{Strateva+05}. The most X-ray weak source WISEA~J033429 is marked in the plot.}
    \label{fig:alpha-OX}
\end{figure}

We search the archived $f_\mathrm{UV}$ values and measure the $f_\mathrm{X}$ according to the best-fitting baseline models of our sources. The host-galaxy contribution is not subtracted from the UV observations, introducing a level of uncertainty. The flux and the corresponding $\alpha_\mathrm{OX}$ values are summarized in columns (4), (5), and (6) of Table~\ref{tab:optical-uv-xray}. We plot the $\alpha_\mathrm{OX}$ versus the $L_{\SI{2500}{\angstrom}}$ in Figure~\ref{fig:alpha-OX}. Most of our higher-redshift sources show good consistency with the $\alpha_\mathrm{OX}-L_{\SI{2500}{\angstrom}}$ relation of radio-quiet type 1 AGNs in \citet{Strateva+05} except for WISEA~J033429. This object is significantly X-ray weak, with $\Delta\alpha_\mathrm{OX}=\alpha_\mathrm{OX,obs}-\alpha_\mathrm{OX,exp}=-0.50$ from the relation, corresponding to an X-ray weakness $f_\mathrm{weak}=10^{-\Delta\alpha_\mathrm{OX}/0.3838}=20.1$. The X-ray weakness indicates the \Lbol\ of WISEA~J033429 estimated from the X-rays is underestimated, which is consistent with the discrepancy between the X-ray and the optical Eddington ratios ($\log\lambda_\mathrm{Edd,X}=-1.60$; $\log\lambda_\mathrm{Edd,opt}=-0.55$) shown Table~\ref{tab:optical-uv-xray}. However, as this source also shows the strongest soft excess emission relative to the continuum (\sx{1} and \sx{2}), the X-weakness is likely to be intrinsic. The light-bending effect in the relativistic reflection model may have an important role. In this case, most of the primary emission from the primary source is bent back toward the BH, causing an X-ray low-flux state \citep{Ross&Fabian+05}.

\section{Discussion}\label{sec:discussion}

\subsection{Systematic Uncertainties and Selection Effects}\label{subsec:systematic}

Here we discuss the systematic effects of our measurement using the baseline models. We note that the most prominent systematic effect lies in the estimation of \kbol. For a majority of the NLS1s in GW+20, \kbol\ were estimated as the average value obtained by the NLS1 sample in \citet{vasudevan+09}. The average bolometric correction factor for NLS1s in \citet{vasudevan+09} is 96.4. If we apply Equation~\ref{eqn:kbol} to the NLS1s in GW+20, based on their rest-frame $2-10$ keV luminosity, the estimated \kbol\ should be $\sim15-30$. Thus, \leddrat\ in GW+20 should be overestimated by a factor of $\sim3-6$, corresponding to a difference of $0.5-0.8\,\rm dex$ for $\log \leddrat$. If we consider this effect in the estimation of \leddrat, the separation between the two samples can naturally be resolved. We test this hypothesis by applying $\kbol=96.4$ as in GW+20 to our NLS1s at higher redshift. We plot the results in purple in Figure~\ref{fig:gamma-mdot}. Our NLS1s at higher redshift now have similar \leddrat\ and $\Gamma$ with the GW+20 NLS1s at lower redshift. This suggests the estimation of \kbol\ can lead to large systematic uncertainties. We emphasize that it does not indicate the estimation of \kbol\ in \citet{netzer+19} is inconsistent with the one in \citet{vasudevan+09}. Rather, the discrepancy in \leddrat\ is because a constant \kbol\ may only work for AGNs with a relatively low \leddrat\ \citep{vasudevan+09}. As our sources show large \leddrat\ ($>0.1$), we prefer an increasing \kbol\ as the measured X-ray luminosities increase, as shown in Equation~\ref{eqn:kbol}. GW+20 did not consider the dependence of \kbol\ on \leddrat, and they did not exclude the outliers in the NLS1 samples of \citet{vasudevan+09}, which can lead to a highly overestimated average \kbol.

Another systematic effect results from the instrumental selection. Our sources have higher redshift, so they need higher absolute luminosity to be detected. On the other hand, they also show larger BH masses, which means they can be more luminous even with a similar \leddrat. This is consistent with the results in Figure~\ref{fig:gamma-mdot} after considering the systematic effect on \kbol. 

A discrepancy in the methods between this work and GW+20 is that GW+20 applied the bulk-motion Comptonization model \texttt{bmc} to model the continuum \citep{Shrader+99_bmc}, rather than the \texttt{powerlaw} model we use here. The \texttt{bmc} model includes four parameters: the seed photon temperature $kT_\mathrm{seed}$, the spectral index $\alpha$, the Comptonization parameter $A$, and the normalization. We test the \texttt{bmc} model by fixing $kT_\mathrm{seed}=0.01$ keV and allow other parameters to vary. The \texttt{bmc} model provides almost the same best-fit results, and the difference of $\Gamma$ compared with our baseline models is less than 0.02, much smaller than the discrepancy detected in Figure~\ref{fig:gamma-mdot}. We thus consider the discrepancy in the baseline models to be negligible.

\phantom{So far we have applied a phenomenological model to quantify the soft excess. Now we want to use more physical models to discuss the origin of the soft excess.}

So far we find that all our sources have significant soft excess despite that they have slightly lower or similar \leddrat\ at higher redshift. We also apply a phenomenological model to quantify the soft excess. Now we want to use more physical models to discuss the origin of the soft excess.

\subsection{The Warm Corona Model}\label{subsec:warm-corona}

In this section, we discuss the possibility of the warm corona model as the origin of the soft excess emission of our sources \citep{Magdziarz+98,Petrucci+18}. Under this hypothesis, the seed UV photons are up-scattered by a warm and optically thick corona. The soft excess is the high-energy tail of the Comptonized component. To do so, we replace the \texttt{bbody} component in our baseline models with the \texttt{nthcomp} model \citep{1999MNRAS.309..561Z} to provide a more physical description of the soft excess. The \texttt{nthcomp} model describes the thermally Comptonized continuum produced in the warm corona. The model considers disk multi-temperature blackbody emission with seed photon temperature at $kT_\mathrm{bb}$ as the seed spectrum. We fix $kT_\mathrm{bb}=10\,\rm eV$ as we only analyze the spectra in the X-ray band. The produced continuum can be easily controlled by the warm corona temperature $kT_\mathrm{e,w}$ and the asymptotic powerlaw photon index $\Gamma_\mathrm{w}$. The powerlaw continuum is also replaced by the \texttt{cutoffpl} model with the high-energy cutoff fixed at 300 keV (we replace the \texttt{powerlaw} model with the \texttt{cutoffpl} model because we want to have a consistent model setup with Section~\ref{subsec:reflection}; see the description therein). As $N_\mathrm{H}$ is unconstrained in our baseline models, we fix the $N_\mathrm{H}$ to Galactic values. The full model in \texttt{Xspec} is
\begin{equation*}
    \texttt{const$\times$TBabs$\times$zashift$\times$(nthcomp+cutoffpl)}.
\end{equation*}

For PG~1543+489, we still add a \texttt{relline} component to model the iron K$\alpha$ line feature. We assume a broken-powerlaw emissivity profile with variable $q_\mathrm{in}$ and $R_\mathrm{br}$ with $\epsilon(r)\propto\epsilon^{-3}$ at $r>R_\mathrm{br}$, which is the same as the treatment in the baseline model. We fix the line energy in the rest frame at $E_\mathrm{line}=6.7\,\rm keV$. Other free parameters in the \texttt{relline} model are the spin ($a_*$), and the inclination angle. The inner radius of the disk is assumed to reach the innermost stable circular orbit ($r_\mathrm{ISCO}$).

\begin{figure*}
    \centering
    \includegraphics[width=\textwidth]{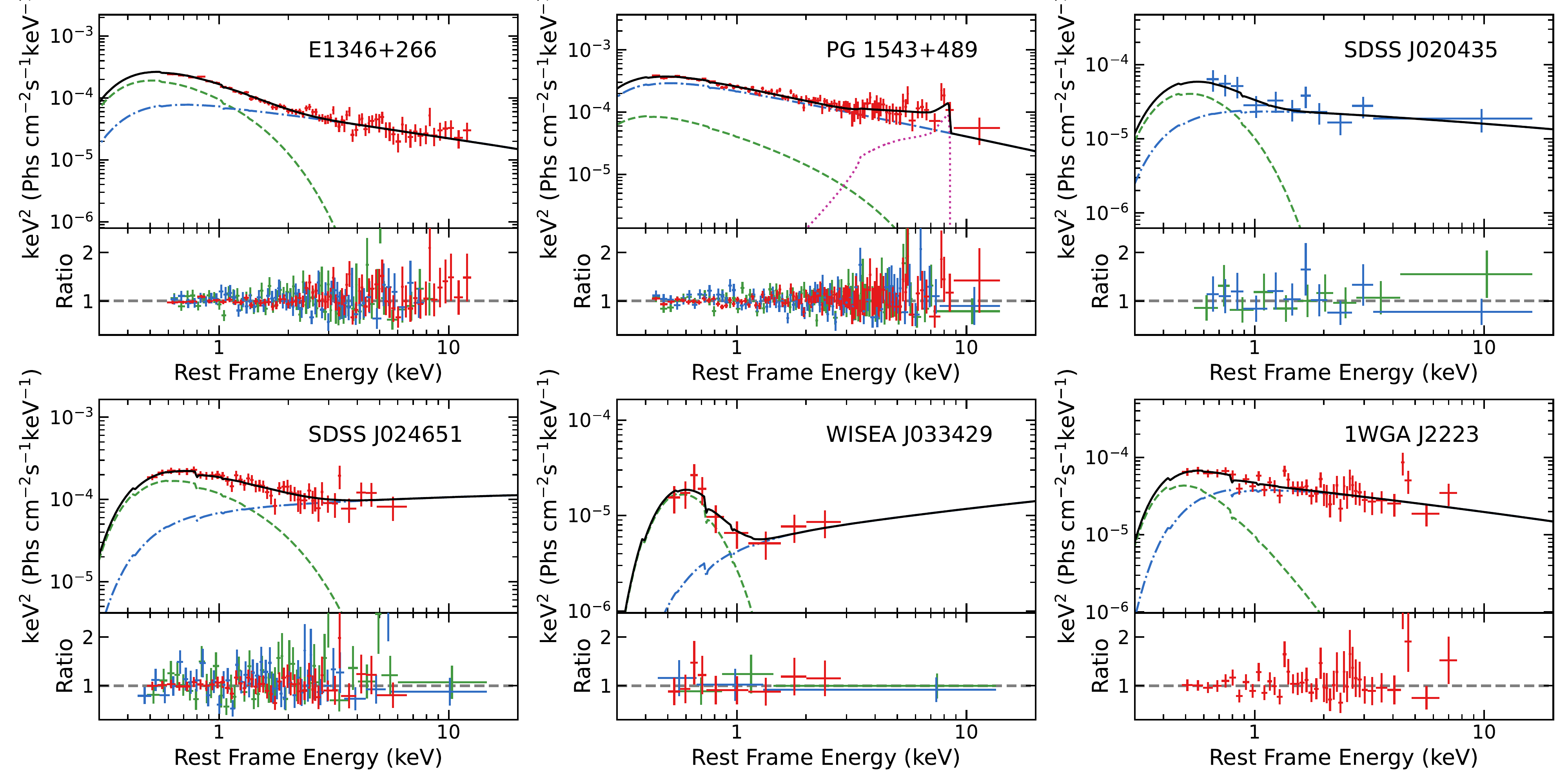}
    \caption{Warm corona model fits. Unfolded best-fit model plots and data-to-model ratio plots for each NLS1. The upper plots show the best-fit models. We only show pn data in red for clarity (MOS1 data in blue for SDSS J020435). Solid black lines: the total model; blue dash-dotted lines: \texttt{cutoffpl}; green dashed lines: \texttt{nthcomp}; purple dotted lines: \texttt{relline}. The lower plots show the data-to-model ratios. The red, blue, and green data points correspond to pn, MOS1, and MOS2 data.}\label{fig:nthcomp}
\end{figure*}

Figure~\ref{fig:nthcomp} shows the best-fit models and data/model ratio plots for the warm corona scenario. The best-fit results are summarized in Table~\ref{tab:nthcomp}. The \texttt{relline} component indicates a fast-spinning BH and a steep powerlaw emissivity profile, consistent with the results where the \texttt{bbody} model is used for the soft excess.

\begin{table*}
\footnotesize
\caption{Best-fit table of the warm corona models. The flux $F$ is in units \fluxcgs. The reported uncertainties correspond to the 90\% confidence level for one parameter ($\Delta\chi^2=2.71$). * indicates that the parameter is frozen in the fit. $\Delta$ means the parameter cannot be constrained.}
\label{tab:nthcomp}
\begin{tabular}{lcccccc}
\hline
\hline
& E1346+266 & PG1543+489 & SDSS J0204 & SDSS J0246 & WISEA J033 & 1WGA J2223\\
\hline
\texttt{nthcomp} &  &  & & & & \\
$\Gamma_\mathrm{w}$ & $3.0_{-1.3}^{+0.6}$ & $3.1_{-0.4}^{+0.8}$ & $<7$ & $2.9_{-1.2}^{+0.4}$ & $2.1\pm\Delta$ & $5.0_{-4}^{+0.6}$\\
$kT_\mathrm{e,w}$ ($\rm keV$) & $>0.17$ & $0.83\pm\Delta$ & $>1.1$ & $>0.16$ & $<1$ & $>0.06$\\
$\log F^\mathrm{nth}_\mathrm{0.3-10\,keV}$ & $-11.9_{-0.3}^{+0.2}$ & $-12.5_{-0.9}^{+0.6}$ & $-12.7_{-0.5}^{+0.5}$ & $-12.0_{-0.4}^{+0.2}$ & $-12.8_{-0.3}^{+0.5}$ & $-12.3_{-0.6}^{+0.5}$\\
\hline
\texttt{cutoffpl} &  &  & & & & \\
$\Gamma_\mathrm{h}$ & $2.5_{-1.3}^{+0.2}$ & $2.7_{-0.4}^{+0.7}$ & $2.2_{-2.0}^{+0.4}$ & $1.9_{-0.8}^{+0.5}$ & $1.7_{-1.0}^{+0.8}$ & $2.4_{-0.4}^{+0.3}$\\
$\log F^\mathrm{pl}_\mathrm{0.3-10\,keV}$ & $-12.1_{-0.9}^{+0.2}$ & $<-11.7$ & $-12.7_{-0.4}^{+0.1}$ & $-12.1_{-0.4}^{+0.2}$ & $-13.3_{-0.1}^{+0.2}$ & $-12.5_{-0.2}^{+0.1}$\\
\hline
\texttt{relline} &  &  & & & & \\
$E_\mathrm{line}$ (keV) &  & $6.7^{*}$ & & & & \\
$q_\mathrm{in}$ &  & $>7$ & & & & \\
$q_\mathrm{out}$ &  & $3^*$ & & & & \\
$R_\mathrm{br}$ ($r_\mathrm{ISCO}$) &  & $1.7_{-0.2}^{+0.8}$ & & & & \\
$a_\mathrm{*}$ &  & $>0.97$ & & & & \\
Incl (deg) &  & $61_{-9}^{+14}$ & & & &\\
$\log F^\mathrm{relline}_\mathrm{0.3-10\,keV}$ & & $-13.0_{-0.3}^{+0.2}$ & & & & \\
\hline
$\chi^2/d.o.f.$ & 164.82/178 & 252.69/237 & 8.22/15 & 113.83/125 & 3.29/8 & 37.73/38\\
\hline
\hline
\end{tabular}
\end{table*}


Our sources present a highly scattered X-ray continuum ($\Gamma_\mathrm{h}=1.7-2.7$) and a warm corona emission with very low temperature and large photon index ($kT_\mathrm{e,w}<0.9\,\mathrm{keV}$, $\Gamma_\mathrm{w}>2.1$). This is consistent with the sample study in \citet{Petrucci+18} where 100 observations of 22 objects were studied (see Figure~5 therein).

\begin{figure}
    \centering
    \includegraphics[width=\columnwidth]{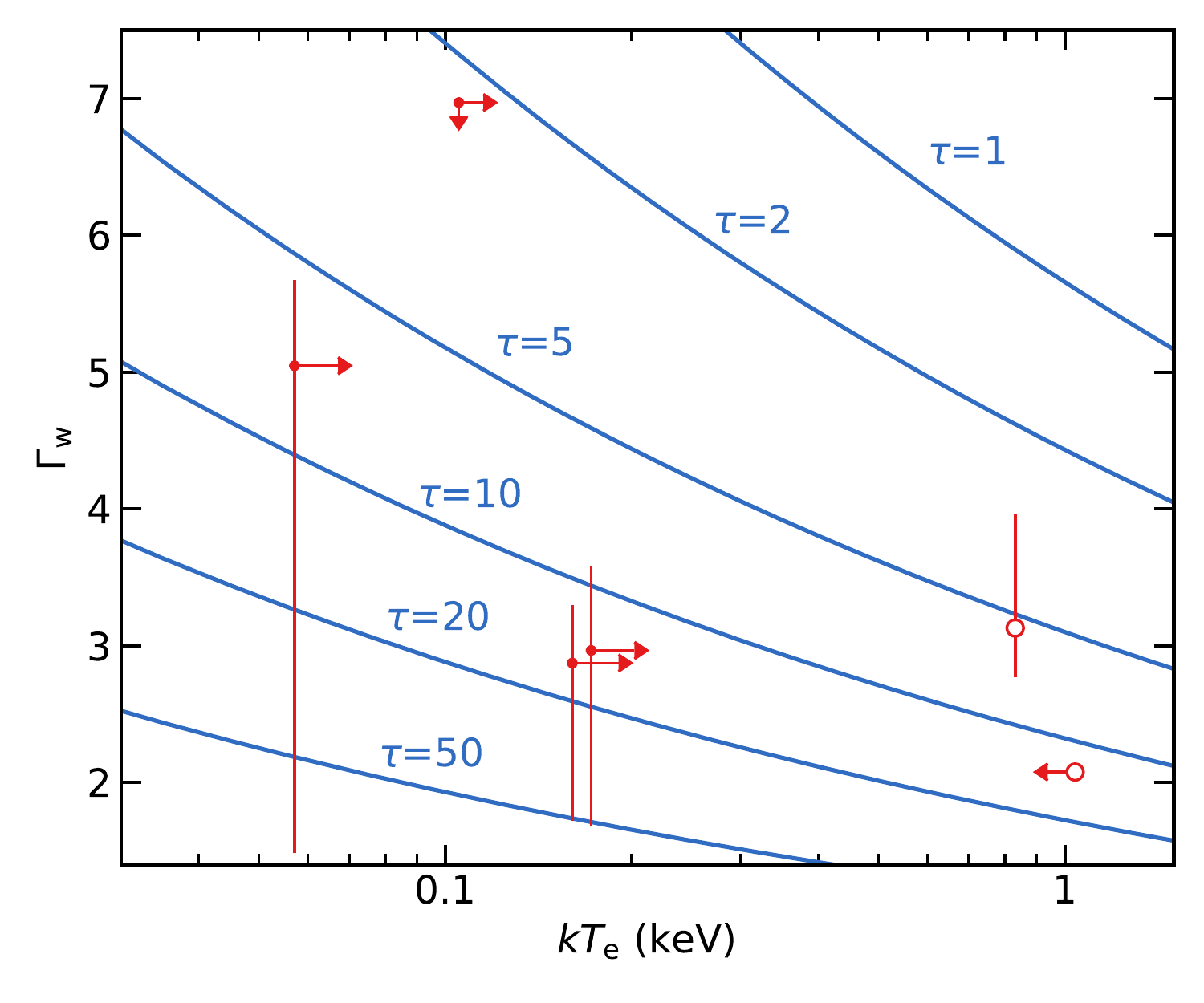}
    \caption{Best-fit warm corona photon index $\Gamma_\mathrm{w}$ versus temperature $kT_\mathrm{e}$ (red points). The solid blue curves are the contours of the optical depth $\tau$ based on Equation~(13) in \citet{1999ASPC..161..295B}. Open circles mean either the temperature or the optical depth of the warm corona is unconstrained.}\label{fig:tau}
\end{figure}

We visualize the warm corona photon index $\Gamma_\mathrm{w}$ and the warm corona temperature $kT_\mathrm{e,w}$ in Figure~\ref{fig:tau}, and the contours of optical depth $\tau$ are also plotted according to Equation~(13) in \citet{1999ASPC..161..295B}. Open circles mean either the temperature or the optical depth of the warm corona is unconstrained. We see that all sources show large $\tau>5$ except for SDSS~J020435, which has an upper limit of $\Gamma_\mathrm{h}$ and a lower limit of $kT_\mathrm{e,w}$. This is in agreement with the previous studies of the warm corona in AGNs \citep[e.g.][]{ 2012MNRAS.420.1825J, 2013A&A...549A..73P, Petrucci+18}.

\subsection{The Relativistic Disk Reflection Model}\label{subsec:reflection}

Another model commonly applied to explain the soft excess is the relativistic reflection model \citep{Crummy+06,Fabian+09,Jiang+18}. The X-ray fluorescence emission is produced by the reprocessing of photons by the inner part of the disk. These fluorescence features will be blurred by the strong relativistic effects around the BH. This effect makes X-ray reflection spectroscopy one of the most effective tools in probing the vicinity of BH. Moreover, the existence of high-density inner-disk electrons ($n_\mathrm{e}>10^{15}\,\mathrm{cm^{-3}}$) would further contribute to an increased emission in the soft X-ray band \citep{Ross&Fabian+93,Garcia+16}.

Currently, the most advanced reflection model package is \texttt{relxill} \citep{2013MNRAS.430.1694D,2014ApJ...782...76G}. It incorporates \texttt{xillver}, which calculates the reflection in the rest-frame of the accretion disk, and the relativistic convolution model \texttt{relconv}. Specifically, to model the reflection component we apply the \texttt{relxillD} flavor which considers a variable disk electron density ($10^{15}<n_\mathrm{e}<10^{19}\ \mathrm{cm^{-3}}$) and a fixed high-energy cutoff at 300 keV. The primary effect of the high density is the increase of free-free heating in the deeper region of the disk, causing an increased disk temperature. The emission of the reflected spectrum at low energies would thus increase and may consequently resolve the existence of the soft excess \citep{Garcia+16,2019MNRAS.489.3436J}. The continuum is described by \texttt{cutoffpl}. The high-energy cutoff is also fixed at 300 keV to keep a consistent setup with the \texttt{relxillD} model (although the exact cutoff value is not important with our data up to $10\,\rm keV$ in the observed frame). We fix the $N_\mathrm{H}$ to Galactic values as we did in the previous analysis. The full model in \texttt{Xspec} notation is
\begin{equation*}
    \texttt{const$\times$TBabs$\times$zashift$\times$(cutoffpl+relxillD)}.
\end{equation*}

Since the hard continuum is treated as the incident spectrum of the reflection component, we link the photon index of \texttt{relxillD} with the spectral slope $\Gamma_\mathrm{h}$ of \texttt{cutoffpl}. For all observations, we assume a powerlaw-like disk emissivity profile $\epsilon(r)\propto r^{-q}$. The inner radius of the disk is assumed to reach the ISCO. In a nominal fit, we allow a variable spin ($a_*$), inclination angle, disk ionization ($\xi$), iron abundance ($A_\mathrm{Fe}$), and disk density ($n_\mathrm{e}$). If neither the upper nor lower limit of $n_\mathrm{e}$ can be constrained, we fix $n_\mathrm{e}=10^{15}\,\rm cm^{-3}$, as in the case of the standard \texttt{relxill} flavor. If the spin cannot be constrained, we fix $a_*=0.998$. We also check that allowing a free inner disk radius $R_\mathrm{in}$ would only result in a $\Delta\chi^2\approx-0.4$ compared with a fixed $R_\mathrm{in}=R_\mathrm{ISCO}$. Thus, we choose to fix the $R_\mathrm{in}$ at the ISCO. If there are other unconstrained parameters, we would further decrease the degrees of freedom by fixing $q=3$, $\theta=45^{\circ}$, and $A_\mathrm{Fe}=1$. In the end, for the least constrained fit, the only free parameters are the photon index $\Gamma_\mathrm{h}$ and the ionization parameter $\xi$, in addition to the normalization.

\begin{figure*}
    \centering
    \includegraphics[width=\textwidth]{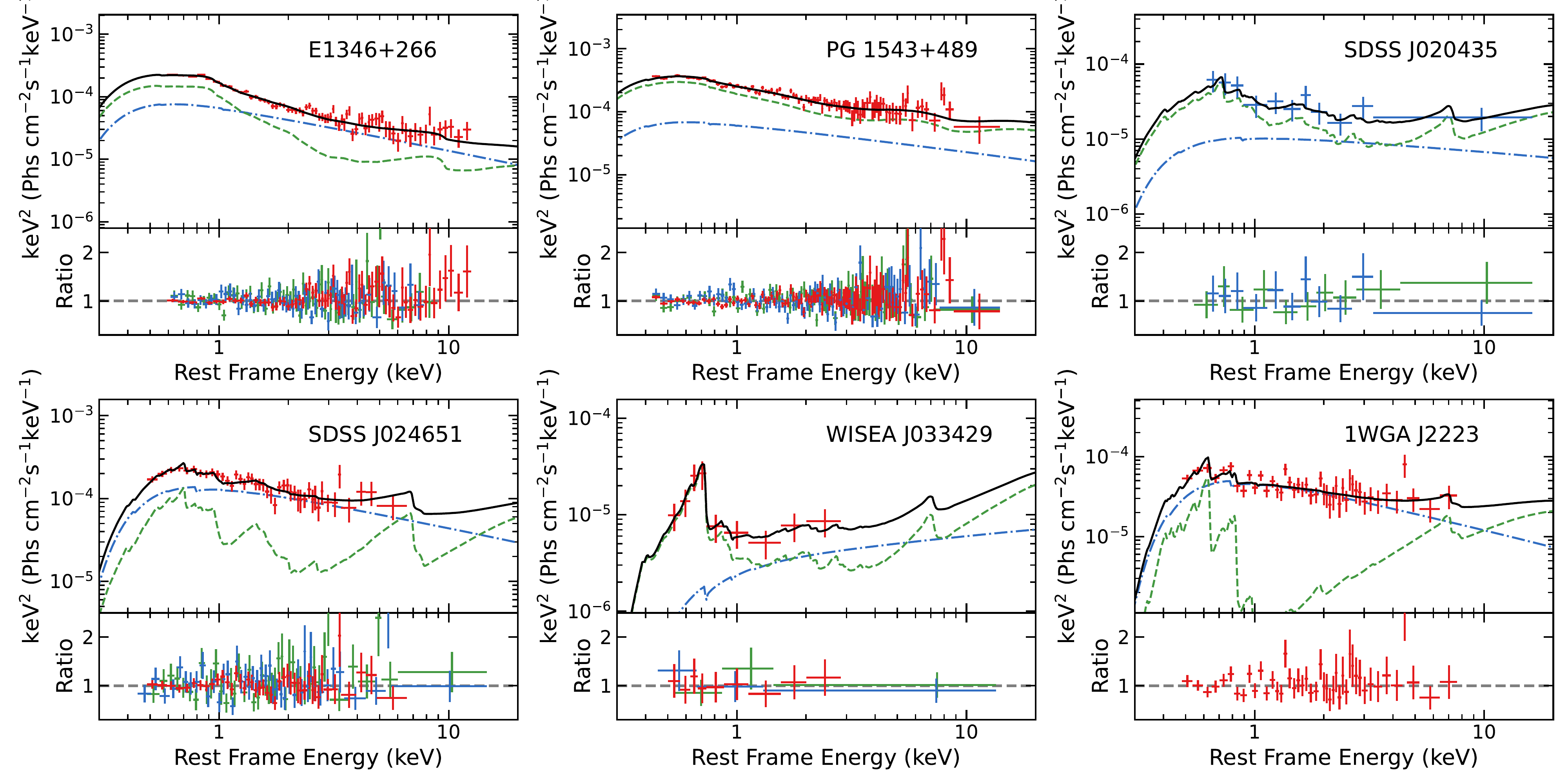}
    \caption{Relativistic reflection model fits. Unfolded best-fit model plots and data-to-model ratio plots for each high-redshift NLS1. The upper plots show the best-fit models. We only show pn data in red for clarity (MOS1 data in blue for SDSS J020435). Solid black lines: the total model; blue dash-dotted lines: \texttt{cutoffpl}; green dashed lines: \texttt{relxillD}. The lower plots show the data-to-model ratios. The red, blue, and green data points correspond to pn, MOS1, and MOS2 data.}\label{fig:relxillD}
\end{figure*}

Figure~\ref{fig:relxillD} shows the best-fit models and data/model ratio plots for the relativistic reflection model. Table~\ref{tab:relxillD} summarizes the best-fit parameter values. The relativistic reflection model provides similarly good fits compared with the warm corona model. All sources show a very soft continuum with $\Gamma_\mathrm{h}>2.2$ except for WISEA~J033429. This source shows the most significant X-ray weakness, and it has a large uncertainty on $\Gamma_\mathrm{h}$ with an upper limit value of 2.4 within 90\% confidence level. The co-existence of the X-ray weakness and the strong \sx{1} and \sx{2} in this source can be explained by the relativistic reflection. The relativistic reflection blurs a series of emission lines in the soft X-ray band, causing an ``excess" of emission; in the meantime, the light-bending effect causes a portion of the primary emission from the hot corona to bend towards the BH and thus results in an X-ray low-flux state. Our spectral fitting confirms a reflection-dominated spectrum in WISEA~J033429 (Figure~\ref{fig:relxillD}), which further supports the relativistic reflection. The conclusion is similar to the one in \citet{Garcia+19}, in which they studied a nearby ($z=0.0344$) bright Seyfert~1 AGN Mrk~509.

\begin{table*}
\footnotesize
\caption{Best-fit table of relativistic reflection models. The ionization parameter $\xi$ is in unit $\rm erg\,cm\,s^{-1}$. The disk electron density $n_\mathrm{e}$ is in units $\rm cm^{-3}$. The flux $F$ is in units \fluxcgs. The disk electron density $n_\mathrm{e}$ is in units $\rm cm^{-3}$. The reported uncertainties correspond to the 90\% confidence level for one parameter ($\Delta\chi^2=2.71$). * indicates that the parameter is frozen in the fit.}
\label{tab:relxillD}
\begin{tabular}{lcccccc}
\hline
\hline
 & E1346+266 & PG1543+489 & SDSS J0204 & SDSS J0246 & WISEA J033 & 1WGA J2223\\
\hline
\texttt{cutoffpl} &  &  & & & & \\
$\Gamma_\mathrm{h}$ & $2.70_{-0.12}^{+0.10}$ & $2.44_{-0.07}^{+0.06}$ & $2.2_{-0.3}^{+0.8}$ & $2.5_{-0.3}^{+0.1}$ & $1.7_{-0.3}^{+0.7}$ & $2.69_{-0.17}^{+0.16}$\\
$\log F^\mathrm{pl}_\mathrm{0.3-10\,keV}$ & $-12.2_{-0.2}^{+0.2}$ & $<-12.0$ & $<-12.5$ & $-12.0_{-0.8}^{+0.1}$ & $<-13.2$ & $-12.39_{-0.06}^{+0.04}$\\
\hline
\texttt{relxillD} &  &  & & & & \\
$q$ & $>6.2$ & $>3$ & $3^*$ & $>0.2$ & $3^*$ & $3^*$\\
$a_\mathrm{*}$ & $>0.95$ & $>0.81$ & $0.998^*$ & $0.998^{*}$ & $0.998^*$ & $0.998^*$\\
Incl (deg) & $82_{-16}^{+4}$ & $69_{-57}^{+4}$ & $45^*$ & $39_{-25}^{+24}$ & $45^*$ & $45^*$\\
$\log \xi$ & $3.0_{-0.2}^{+0.4}$ & $3.3_{-1.4}^{+0.4}$ & $<3.5$ & $2.0_{-0.6}^{+1.1}$ & $2.3_{-0.7}^{+0.9}$ & $<1.4$\\
$A_\mathrm{Fe}\,(A_\mathrm{Fe,\odot})$ & $>1.1$ & $0.9_{-0.2}^{+1.5}$ & $1^*$ & $>3.2$ & $1^*$ & $1^*$\\
$\log n_\mathrm{e}$ & $17.4_{-0.9}^{+1.1}$ & $<18$ & $15^*$ & $<18$ & $>17.9$ & $15^*$\\
$\log F^\mathrm{relxill}_\mathrm{0.3-10\,keV}$ & $-11.99_{-0.19}^{+0.19}$ & $-11.8_{-0.4}^{+0.2}$ & $-12.6_{-0.8}^{+0.2}$ & $-12.3_{-0.3}^{+0.4}$ & $-13.0_{-0.3}^{+0.4}$ & $-12.9_{-0.2}^{+0.1}$ \\
\hline
$\chi^2/d.o.f.$ & 163.83/174 & 254.95/239 & 8.37/16 & 107.09/121 & 2.71/8 & 37.61/39\\
\hline
\hline
\end{tabular}
\end{table*}

Interestingly, our sources show slightly higher ionization than typical Sy1s \citep[$\log\xi=1-2$, see e.g.,][]{Walton+13}. A high ionization can contribute to the soft excess together with a high-density disk. The BH spins of these sources are also not well constrained due to a low SNR in the iron band. The most stringent constraints come from E1346+266 and PG~1543+489. Both sources indicate a rapidly spinning BH ($a_*>0.94$ and $a_*>0.81$, respectively).

\begin{figure}
    \centering
    \includegraphics[width=\columnwidth]{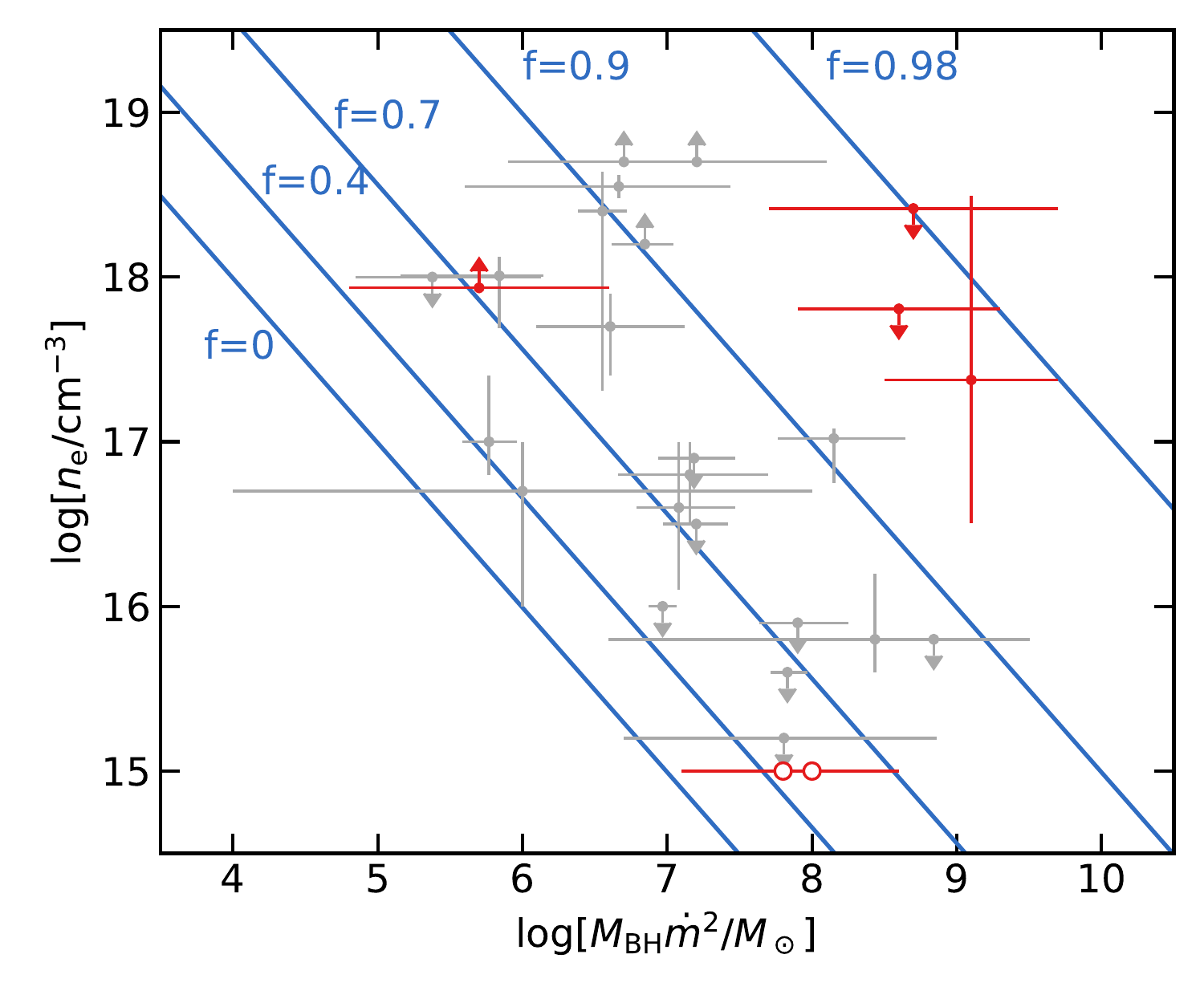}
    \caption{Disk electron density $\log n_\mathrm{e}$ versus $\log(M_\mathrm{BH}\dot{m}^2)$. The grey points are the AGN samples considered in \citet{2019MNRAS.489.3436J}, and the red points
    are the high-z NLS1s in our sample. The solid blue lines are the solutions of disk density in Equation~(8) in \citet{1994ApJ...436..599S} at $r=2R_\mathrm{s}$ for $f=0$ and $f=0.9$, assuming $R_\mathrm{in}=R_\mathrm{ISCO}=1.24R_\mathrm{g}$ with $\xi^{'}=1$, $\alpha=0.1$, $\eta=0.19\pm0.13$ for a maximally spinning Kerr black hole. The open circles indicate that the results cannot be constrained.}\label{fig:ne-mmdot}
\end{figure}

According to the standard $\alpha$-disk model by \citet{1973A&A....24..337S}, \citet{1994ApJ...436..599S} gave a relationship between the density of a radiation-pressure-dominated disk at radius $r$ and the accretion rate
\begin{equation}
     n_{\mathrm{e}}(r)=\frac{1}{\sigma_{\mathrm{T}} R_{\mathrm{S}}} \frac{256 \sqrt{2}}{27} \alpha^{-1} r^{3 / 2} \dot{m}^{-2}\left[1-\left(R_{\mathrm{in}} / r\right)^{1 / 2}\right]^{-2}[\xi^{'}(1-f)]^{-3},\label{eqn:ne}
\end{equation}
where $\sigma_\mathrm{T}$ is the Thompson cross section, $R_\mathrm{S}\equiv2GM_\mathrm{BH}/c^2$ is the Schwarzchild radius, and $r$ is the radius on the disk in units of $R_\mathrm{S}$. We take $\alpha=0.1$ as the standard viscosity parameter \citep{1973A&A....24..337S}, and $\xi^{'}=1$ in the radiative diffusion equation, and $f$ is the fraction of the total transported accretion power released from the disk to the hot corona \citep{2021ApJ...913...13X}. The dimensionless accretion rate $\dot{m}$ is defined as $\dot{m}\equiv\dot{M}/\dot{M}_\mathrm{crit}=L/\eta\ledd$, where $\eta$ is the efficiency for converting accreted mass into outgoing radiation. $\eta=0.057$ for a Schwarzschild BH ($a_*=0$) and $\eta=0.32$ for a maximally spinning Kerr BH ($a_*=0.998$) \citep{1974ApJ...191..507T}. We assume $\eta=0.19\pm0.13$ in our analysis so that it takes into account all possible values of $\eta$, and we propagate the uncertainties to the estimated accretion rates of our sample. According to Equation~\ref{eqn:ne}, $\log n_\mathrm{e}\propto-\log(M_\mathrm{BH}\dot{m}^2)$. Figure~\ref{fig:ne-mmdot} shows $\log n_\mathrm{e}$ versus $\log(M_\mathrm{BH}\dot{m}^2)$. The solid lines show the analytic solution for different values of $f$ assuming $r=2R_\mathrm{s}$. The red points are the high-$z$ sources in our sample, and the grey points are the samples studied in \citet{2019MNRAS.489.3436J}. The open circles indicate that the results cannot be constrained. Taking into account the uncertainties in $\log n_\mathrm{e}$, our sources roughly follow the trend as predicted in \citet{1994ApJ...436..599S} that a system with lower $\log M_\mathrm{BH}$ tends to have higher $n_\mathrm{e}$. The only source with constrained measurements of $n_\mathrm{e}$ is E1346+266, which has the largest mass and highest redshift among our sources. In this source, about 95\% of the energy in the disk is released to the corona, and the radiation-pressure dominated region is $r<2R_\mathrm{s}$ to explain the inferred high disk density.

\section{Conclusion}\label{sec:conclusion}

We performed a detailed spectral analysis of six NLS1s at $z=0.35-0.92$ using \xmm\ EPIC data. Our samples show large BH masses at the upper end of the local samples ($\log\,\mbh>7.5$) and slightly lower or similar \leddrat. We discover that all these sources have strong soft excess emission below 2 keV with relatively soft continuum.

After confirming the existence of the soft excess emission, we use the phenomenological \texttt{bbody} model to quantify the soft excess strength and compare the results from our higher-redshift sample with the local NLS1s and BLS1s in GW+20. We find that the soft excess strength does not show much deviation from the previous sample. We also obtain the $\alpha_\mathrm{OX}-L_\mathrm{\SI{2500}{\angstrom}}$ relation and find the most X-ray-weak source WISEA~J033429 is intrinsically X-ray weak judging from its strong \sx{1} and \sx{2}. We discuss in detail the impact of the bolometric correction factor \kbol, and conclude it carries the major systematic uncertainties in our measurements. The systematic difference of the estimated \kbol\ can explain the discrepancy of \leddrat\ between our high-redshift sample and GW+20 sample. This further suggests the higher absolute luminosities of our sources are driven by BH mass rather than \leddrat. For such detailed spectral analysis in our work, the high-redshift and the soft nature of the sources pose a great challenge for us to accurately identify the continuum, which results in a degeneracy in the parameters.

We then further consider more physical scenarios, the warm-corona model and the relativistic reflection model, to account for the soft excess. These two models provide almost equally good fits. They also allow us to probe the properties of the innermost region of the AGNs. We show that in the warm-corona scenario, the warm coronae of our sample have a low temperature ($kT_\mathrm{e,w}\approx0.1-1\ \rm keV$) and large optical depth ($\tau=5-20$); in the reflection scenario, the sources show relatively higher ionization than the samples in \citet{Walton+13}. We are also able to constrain the high spin values of several sources. Combining both facts that WISEA~J033429 shows the strongest relative soft excess strength with respect to the continuum emission and it shows a reflection-dominated spectrum, we tentatively conclude the relativistic reflection is a more promising explanation for the soft excess, though we cannot exclude the warm corona model based on the statistics. in the entire. Detailed tests of different models for the soft excess are better carried out with bright local samples, which is beyond the scope of this work.

The low SNR in the iron line band does not allow us to identify the iron line feature confidently. Observations from future X-ray missions like \textit{Athena} may be able to provide better constraints on the spectral shape. The large effective area and energy resolution in the iron line band would allow us to better model the reflection spectrum and might help determine the contribution to the soft excess \citep{Parker+22_AGN-degen}. 

While both the warm-corona and relativistic reflection are possible explanations for the soft excess, it is natural to consider reflection from the disk in addition to Comptonization in a warm corona \citep{2021ApJ...913...13X}. Recently, \citet{2022arXiv220606825X} used a new model \texttt{reXcor} that self-consistently combines a warm corona and ionized reflection, and analyzes two Seyfert galaxies HE~1131-1820 and NGC~4593. Although the geometry of the systems is much simplified, the model was able to produce several broadened reflection features with a variety of shapes of the soft excess. Such a type of model may provide new insights about the NLS1s at higher redshift.

\section*{Acknowledgements}

Z.Y. thanks Mario Gliozzi for sharing the data in GW+20 and acknowledges support from the Verne M. Willaman Distinguished Graduate Fellowships in Science. J.J. acknowledges support from the Leverhulme Trust, the Isaac Newton Trust, and St Edmund's College.

\section*{Data Availability}
 
The data underlying this article are available in the High Energy Astrophysics Science Archive Research Center (HEASARC), at \url{https://heasarc.gsfc.nasa.gov}. The \texttt{relxill} package can be downloaded at \url{http://www.sternwarte.uni-erlangen.de/~dauser/research/relxill/}.



\bibliographystyle{mnras}
\bibliography{bibtex} 



\appendix

\section{Candidate NLS1 Sample}\label{app:candidates}

Here we present a list of candidate sources we have examined before our analysis in Table~\ref{tab:candidate}. These sources are identified as NLS1s in the optical bands. We examine their X-ray visibility in the \xmm\ archive. We include the six sources with the highest count rates ($>0.020\,\rm cts/s$) in our analysis.

\begin{table}
\caption{Candidate NLS1s for our analysis. The selected NLS1s are marked in boldface.}
\label{tab:candidate}
\begin{tabular}{lccc}
\hline
 Full Name & $z$ & \xmm & EPIC-pn count rate\\
 & & ObsID & upper limit (cts/s)\\
\hline
\textbf{E1346+266} & 0.92 & 0109070201 & 0.310\\
\textbf{PG~1543+489} & 0.40 & 0153220401 & \\
 & & 0505050201 & 0.512 (stacked)\\
 & & 0505050701 & \\
SDSS J014123.32$+$000752.8 & 0.97 & 	
0747430101 & 0.019\\
\textbf{SDSS J020435.18$-$093154.9} & 0.62 & 0763910301 & 0.061     \\
\textbf{SDSS J024651.91$-$005930.9} & 0.47 & 0402320101 & 0.369\\
SDSS J150728.65$+$523137.6 & 0.85 & 0804620201 & 0.005\\
\textbf{WISEA J033429.44$+$000610.9} & 0.35 & 0402320201 & 0.021\\
\textbf{1WGA J2223.7$-$0206} & 0.46 & 0090050601 & 0.097\\
\hline
\end{tabular}
\end{table}


\bsp	
\label{lastpage}

\end{document}